\documentclass[12pt,a4paper]{article}
\usepackage[utf8]{inputenc}
\usepackage{amsmath}
\usepackage{amsfonts}
\usepackage{amssymb}
\usepackage{graphicx}
\usepackage{rotating}
\usepackage{pdflscape}
\usepackage{xspace}
\usepackage{xcolor}
\usepackage{cite}
\usepackage{mciteplus}
\usepackage{upgreek}
\usepackage{hyperref}
\usepackage[all]{hypcap}
\usepackage{ifthen}
\usepackage{authblk}

\usepackage{lineno}

\newboolean{uprightparticles}
\setboolean{uprightparticles}{false} 

\newboolean{articletitles}
\setboolean{articletitles}{true}

\def\MagUp {\mbox{\em Mag\kern -0.05em Up}\xspace}

\ifthenelse{\boolean{uprightparticles}}%
{

 \def\Pmu         {\ensuremath{\upmu}\xspace}                 
 \def\Pnu         {\ensuremath{\upnu}\xspace}                 
                  
 \def\Ppi         {\ensuremath{\uppi}\xspace}

 \def\PDelta      {\ensuremath{\Delta}\xspace}                 
 \def\PXi      {\ensuremath{\Xi}\xspace}                 
 \def\PLambda      {\ensuremath{\Lambda}\xspace}                 
 \def\PSigma      {\ensuremath{\Sigma}\xspace}                 
 \def\POmega      {\ensuremath{\Omega}\xspace}                 
 \def\PUpsilon      {\ensuremath{\Upsilon}\xspace}                 
 
 \def\PB      {\ensuremath{\mathrm{B}}\xspace}                 
                  
 \def\PD      {\ensuremath{\mathrm{D}}\xspace}

 \def\PK      {\ensuremath{\mathrm{K}}\xspace}

 \def\Pb      {\ensuremath{\mathrm{b}}\xspace}                 
 \def\Pc      {\ensuremath{\mathrm{c}}\xspace}

 \def\Pi      {\ensuremath{\mathrm{i}}\xspace}

 \def\Pn      {\ensuremath{\mathrm{n}}\xspace}                 
                  
 \def\Pp      {\ensuremath{\mathrm{p}}\xspace}

 \def\Ps      {\ensuremath{\mathrm{s}}\xspace}

}
{

 \def\Pmu         {\ensuremath{\mu}\xspace}                 
 \def\Pnu         {\ensuremath{\nu}\xspace}                 
                  
 \def\Ppi         {\ensuremath{\pi}\xspace}

 \mathchardef\PDelta="7101
 \mathchardef\PXi="7104
 \mathchardef\PLambda="7103
 \mathchardef\PSigma="7106
 \mathchardef\POmega="710A
 \mathchardef\PUpsilon="7107
                  
 \def\PB      {\ensuremath{B}\xspace}                 
                  
 \def\PD      {\ensuremath{D}\xspace}

 \def\PK      {\ensuremath{K}\xspace}

 \def\Pb      {\ensuremath{b}\xspace}                 
 \def\Pc      {\ensuremath{c}\xspace}

 \def\Pi      {\ensuremath{i}\xspace}

 \def\Pn      {\ensuremath{n}\xspace}                 
                  
 \def\Pp      {\ensuremath{p}\xspace}

 \def\Ps      {\ensuremath{s}\xspace}

}

\makeatletter
\ifcase \@ptsize \relax
  \newcommand{\miniscule}{\@setfontsize\miniscule{4}{5}}
\or
  \newcommand{\miniscule}{\@setfontsize\miniscule{5}{6}}
\or
  \newcommand{\miniscule}{\@setfontsize\miniscule{5}{6}}
\fi
\makeatother

\DeclareRobustCommand{\optbar}[1]{\shortstack{{\miniscule (\rule[.5ex]{1.25em}{.18mm})}
  \\ [-.7ex] $#1$}}

\def\mup        {{\ensuremath{\Pmu^+}}\xspace}

\def\neu        {{\ensuremath{\Pnu}}\xspace}

\def\neum       {{\ensuremath{\neu_\mu}}\xspace}

\def\squark    {{\ensuremath{\Ps}}\xspace}

\def\cquark    {{\ensuremath{\Pc}}\xspace}

\def\bquark    {{\ensuremath{\Pb}}\xspace}

\def\pion   {{\ensuremath{\Ppi}}\xspace}

\def\pip    {{\ensuremath{\pion^+}}\xspace}

\def\kaon    {{\ensuremath{\PK}}\xspace}
  \def\Kbar    {{\kern 0.2em\overline{\kern -0.2em \PK}{}}\xspace}

\def\KorKbar    {\kern 0.18em\optbar{\kern -0.18em K}{}\xspace}

\def\Km      {{\ensuremath{\kaon^-}}\xspace}

\def\KL      {{\ensuremath{\kaon^0_{\rm\scriptscriptstyle L}}}\xspace}

  \def\Dbar    {{\kern 0.2em\overline{\kern -0.2em \PD}{}}\xspace}
\def\D       {{\ensuremath{\PD}}\xspace}

\def\DorDbar    {\kern 0.18em\optbar{\kern -0.18em D}{}\xspace}
\def\DtwoorDtwobar {\kern -0.25em\optbar{\kern 0.25em D_2^*}{}\xspace}

\def\Dsm     {{\ensuremath{\D^-_\squark}}\xspace}

\def\B       {{\ensuremath{\PB}}\xspace}
\def\Bbar    {{\ensuremath{\kern 0.18em\overline{\kern -0.18em \PB}{}}}\xspace}

\def\BorBbar    {\kern 0.18em\optbar{\kern -0.18em B}{}\xspace}
\def\Bz      {{\ensuremath{\B^0}}\xspace}

\def\BzorBzbar  {\kern 0.18em\optbar{\kern -0.18em B}{}^0\xspace}

\def\Bs      {{\ensuremath{\B^0_\squark}}\xspace}
\def\Bsb     {{\ensuremath{\Bbar{}^0_\squark}}\xspace}

  \def\Y#1S{\ensuremath{\PUpsilon{(#1S)}}\xspace}

\def\proton      {{\ensuremath{\Pp}}\xspace}

\def\neutron     {{\ensuremath{\Pn}}\xspace}
\def\antineutron {{\ensuremath{\overline \neutron}}\xspace}

\def\Lz          {{\ensuremath{\PLambda}}\xspace}
\def\Lbar        {{\ensuremath{\kern 0.1em\overline{\kern -0.1em\PLambda}}}\xspace}
\def\LorLbar    {\kern 0.18em\optbar{\kern -0.18em \PLambda}{}\xspace}

\def\Lc      {{\ensuremath{\Lz^+_\cquark}}\xspace}

\def\to                 {\ensuremath{\rightarrow}\xspace}

\def\AT#1     {\ensuremath{A_{\mathrm{T}}^{#1}}\xspace}           

\def\C#1      {\ensuremath{\mathcal{C}_{#1}}\xspace}                       
\def\Cp#1     {\ensuremath{\mathcal{C}_{#1}^{'}}\xspace}                    
\def\Ceff#1   {\ensuremath{\mathcal{C}_{#1}^{\mathrm{(eff)}}}\xspace}        
\def\Cpeff#1  {\ensuremath{\mathcal{C}_{#1}^{'\mathrm{(eff)}}}\xspace}       
\def\Ope#1    {\ensuremath{\mathcal{O}_{#1}}\xspace}                       
\def\Opep#1   {\ensuremath{\mathcal{O}_{#1}^{'}}\xspace}                    

\newcommand{\tev}{\ifthenelse{\boolean{inbibliography}}{\ensuremath{~T\kern -0.05em eV}\xspace}{\ensuremath{\mathrm{\,Te\kern -0.1em V}}}\xspace}
\newcommand{\gev}{\ensuremath{\mathrm{\,Ge\kern -0.1em V}}\xspace}
\newcommand{\mev}{\ensuremath{\mathrm{\,Me\kern -0.1em V}}\xspace}
\newcommand{\kev}{\ensuremath{\mathrm{\,ke\kern -0.1em V}}\xspace}
\newcommand{\ev}{\ensuremath{\mathrm{\,e\kern -0.1em V}}\xspace}
\newcommand{\gevc}{\ensuremath{{\mathrm{\,Ge\kern -0.1em V\!/}c}}\xspace}
\newcommand{\mevc}{\ensuremath{{\mathrm{\,Me\kern -0.1em V\!/}c}}\xspace}
\newcommand{\gevcc}{\ensuremath{{\mathrm{\,Ge\kern -0.1em V\!/}c^2}}\xspace}
\newcommand{\gevgevcccc}{\ensuremath{{\mathrm{\,Ge\kern -0.1em V^2\!/}c^4}}\xspace}
\newcommand{\mevcc}{\ensuremath{{\mathrm{\,Me\kern -0.1em V\!/}c^2}}\xspace}

\def\mum  {\ensuremath{{\,\upmu\rm m}}\xspace}

\def\ps   {\ensuremath{{\rm \,ps}}\xspace}

\def\gsim{{~\raise.15em\hbox{$>$}\kern-.85em
          \lower.35em\hbox{$\sim$}~}\xspace}
\def\lsim{{~\raise.15em\hbox{$<$}\kern-.85em
          \lower.35em\hbox{$\sim$}~}\xspace}

\def\pt         {\mbox{$p_{\rm T}$}\xspace}

\def\tell1  {TELL1\xspace}
\def\ukl1   {UKL1\xspace}

\newcommand{\eg}{\mbox{\itshape e.g.}\xspace}
\newcommand{\ie}{\mbox{\itshape i.e.}\xspace}

\title{Oscillations of $B_s^0$ mesons as a probe of decays with unreconstructed particles}
\author{Anton Poluektov}
\author{Adam Morris}
\affil{\small Aix Marseille Univ, CNRS/IN2P3, CPPM, Marseille, France}

\begin{document}


\maketitle

\begin{abstract}
  \noindent
  A new experimental technique to study the decays of \Bs mesons into final states involving unreconstructable 
  particles is proposed.  The method uses a characteristic feature of fast flavour oscillations of $\Bs$ mesons. 
  Analysis of the oscillation-frequency spectrum of partially reconstructed flavour-tagged 
  $\Bs$ decays provides information about the mass spectrum of unreconstructed decay products and 
  suppresses any background that does not come from the decays of \Bs mesons. 
\end{abstract}

\section{Introduction}

\label{sec:introduction}

Many interesting measurements in heavy flavour physics, both within the Standard Model (SM) framework and beyond the SM, involve decays of heavy-flavoured hadrons into final states with neutral particles that are difficult or impossible to reconstruct with collider detectors. The SM decays may involve decays with neutrinos, neutrons, or $\KL$ mesons. In addition, non-SM phenomena may generate final states involving heavy neutrinos, axion-like particles or other dark matter candidates. 

Various experimental techniques have been developed to deal with such decays. In experiments at lepton colliders, one can use kinematic constraints coming from the well-defined initial state.
For instance, at $B$-factories, where pairs of $B$ mesons are produced at threshold without additional particles, full reconstruction of one of the $B$ mesons provides sufficient kinematic constraints to reconstruct undetected particles in the decay of the other $B$ meson \cite{Lees:2012wv,Hsu:2012uh,Lutz:2013ftz}. At the LHCb experiment~\cite{Alves:2008zz}, which uses hadron collisions, techniques employing the information about the topology of the decays have been adopted for decays with neutrinos in the final state 
\cite{Aaij:2015bfa, Stone:2014mza}. These techniques take advantage of the large boost of initial-state hadrons in proton interactions at the LHC as well as the excellent vertex resolution of the LHCb detector to constrain the missing kinematic information by using displaced secondary vertices from weak decays. 

In this paper, we propose a novel technique that uses the known pattern of \Bs oscillations as a tool to study a specific family of \Bs decays with invisible (unreconstructed) particles in the final state. 
The method uses the characteristically high frequency of \Bs flavour oscillations, which leads to multiple changes of flavour on the scale of one \Bs lifetime. We show that, when combined with the information about the topology of the \Bs decay, this feature provides a strong kinematic constraint on decays with invisible particles and efficiently suppresses any non-\Bs backgrounds. Contrary to other methods involving topological reconstruction at LHCb, where assumptions are needed on the nature of the missing particle (\eg, that its mass is equal to zero), this technique provides the information about the mass (or even the spectrum of invariant masses) of the invisible state(s). 

\begin{figure}
  \begin{center}
  \includegraphics[width=0.65\textwidth]{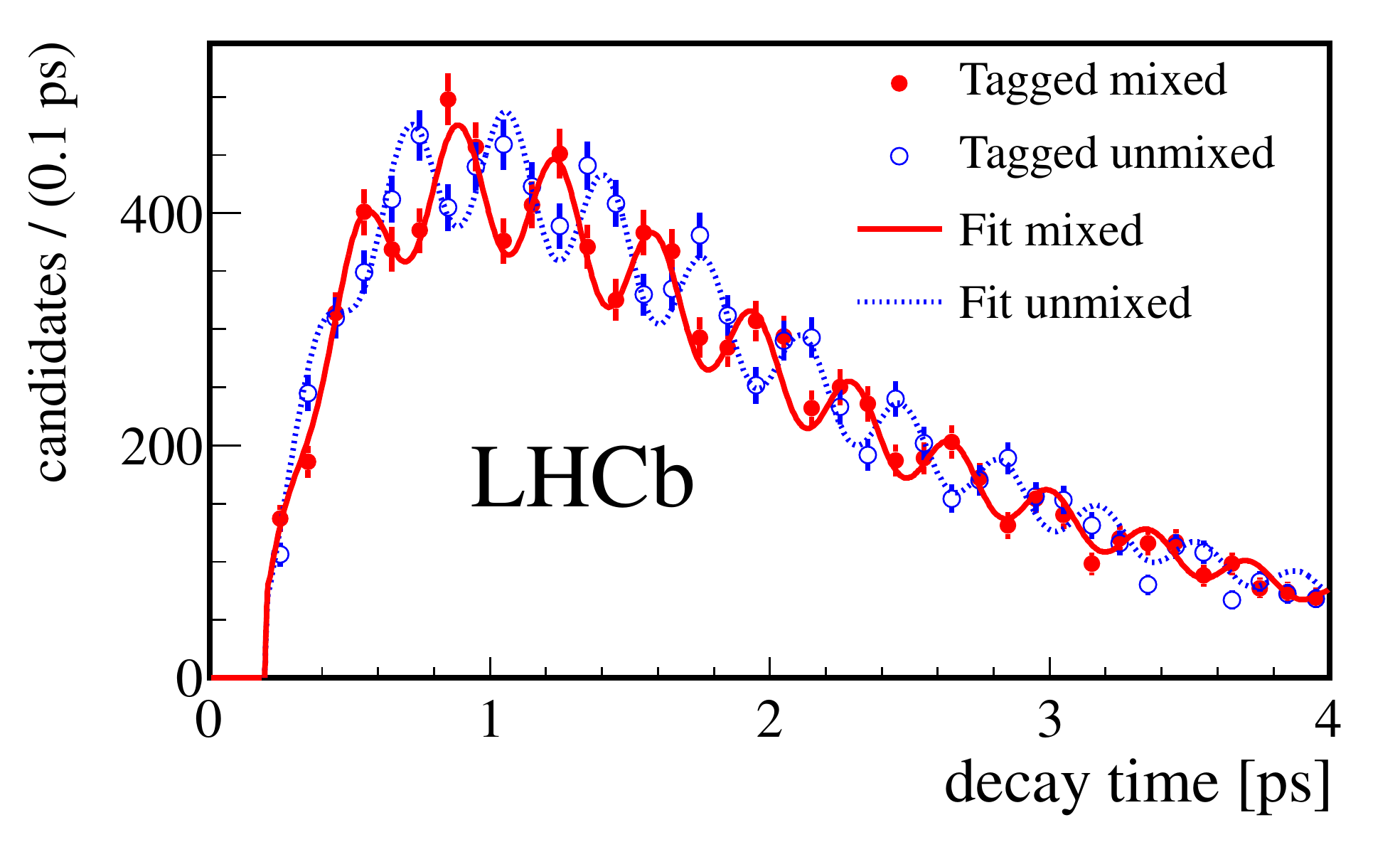}
  \end{center}
  \caption{Decay time distribution of flavour-tagged $\Bs$ mesons reconstructed with the LHCb detector in the $\Bs\to\Dsm\pip$ decays~\cite{LHCb-PAPER-2013-006}. }
  \label{fig:bs_mixing}
\end{figure}

The angular frequency of $\Bs$ meson oscillations $\Delta m_s$ has been precisely measured to be $17.757\pm 0.021\ps^{-1}$, corresponding to the mixing parameter $x_s=\Delta m_s/\Gamma_s = 26.81\pm 0.08$~\cite{PhysRevD.98.030001} (where $\Gamma_s$ is the average $\Bs$ decay width)\footnote{Natural units with $\hbar=c=1$ are used.}. The most precise measurement of $\Delta m_s$
comes from the analysis of $\Bs\to\Dsm\pip$ 
decays\footnote{Charge conjugation is implied throughout this paper.} by the LHCb collaboration~\cite{LHCb-PAPER-2013-006}. The distribution of decay time for mixed (where the 
flavour of the \Bs meson at production and decay are different) and unmixed \Bs meson decays in this analysis is shown in Fig.~\ref{fig:bs_mixing}. 
Observing \Bs oscillations requires knowledge of the flavour of the \Bs meson at the time of production (flavour tagging), as well as precise 
measurement of the decay time. LHCb takes advantage of the large boost of \Bs mesons in the laboratory frame 
to transform the measurement of the flight distance of the \Bs meson to the measurement of its 
decay time. This transformation requires knowledge of the \Bs momentum. 

In our proposed approach, we use the fact that the momentum of the partially reconstructed \Bs meson can be determined from the position of its decay vertex, assuming that the 
mass of the invisible state is known. Similar topological reconstruction has been used in studies of 
semileptonic decays at LHCb~\cite{LHCb-PAPER-2017-027}. With the correctly determined \Bs momentum, the decay time distribution 
will exhibit oscillations with a known frequency. One can thus use the amplitude of oscillations at that frequency to constrain 
the mass (or measure the spectrum of masses) of the invisible particles in partially reconstructed \Bs decays. 

The method is based on the reconstruction of momenta from the decay topology, thus its applicability is critically dependent on the vertex resolution of the detector. Simulation studies show that the technique is feasible with the value of vertex resolution corresponding to one reached at the LHCb experiment. 

Since the method relies on the observation of \Bs oscillations, the family of decays that could be studied must satisfy two conditions: the decay should be flavour-specific (\ie, the decays of \Bs and \Bsb mesons to the final state under consideration should at least have different probabilities, or one of them should be suppressed), and the decay vertex should be well-reconstructable. Possible applications of this
technique could be studies of decays with heavy invisible (or poorly reconstructed) particles in the final state, both for SM decays, \eg, involving neutrons, and for searches for non-SM particles. One particular example of the processes beyond the SM that satisfy the above conditions is the model suggested in Refs.~\cite{Elor:2018twp, Nelson:2019fln, Alonso-Alvarez:2019fym}, where a heavy invisible particle is accompanied by a SM baryon. As a benchmark SM decay mode, one can consider the decay $\Bs\to \Dsm\proton\antineutron$, which satisfies both conditions and should have a significant rate since it proceeds via the favoured $\bquark\to \cquark$ transition.  
Reconstruction of decays with SM neutrinos, such as $\Bs\to\Dsm\mup\neum$, can also be considered; in that case, this technique will provide the additional constraint for the reconstruction of kinematic variables related to neutrinos, such as the invariant mass of the $\mup\neum$ system. 

\section{Formalism}

Consider the decay $\Bs\to AX$, where the system $A$ (it can be a single state or a combination of several particles) is reconstructed, and $X$ is invisible or partially reconstructed. Consider as well that the system
$A$ tags the flavour of $\Bs$ meson, {\ie}, decays of $\Bs$ and $\Bsb$ mesons to the final state $AX$ at least do not occur at the same rate 
or one of them is forbidden. The system $A$ or the combination $AX$ is also required to have a well-reconstructable vertex that coincides with the \Bs decay vertex, \eg, it should be reconstructed from at least two charged particles. 

To be specific, let us give an example in the context of searches for dark matter (DM) candidates in $B$ meson decays suggested in 
Ref.~\cite{Elor:2018twp}. In that model, transitions such as $\bar{\bquark}\to\cquark\squark\psi$ are expected, where $\psi$ is a heavy neutral DM particle with negative baryonic charge. 
This transition gives rise to such \Bs meson decays as $\Bs\to \Xi_{\cquark}^0\psi$, where $\psi$ would act as the invisible state $X$ in the notation used in our formalism. While the ground 
state $\Xi_{\cquark}^{0}$ is long-lived and does not constrain the \Bs decay position, the excited $\Xi_{\cquark}^{*0}$ state decaying to $\Lc\Km$ forms a well-defined vertex and can serve as the state $A$ in our notation. 
Similarly, for the SM decay $\Bs\to\Dsm\proton\antineutron$ mentioned above, the antineutron serves as the invisible particle $X$, while the \Bs decay vertex is provided by the combination $\Dsm\proton$ that acts as the state $A$. In the reconstruction of semileptonic decays such as $\Bs\to\Dsm\mup\neum$, where the neutrino is assumed to be massless, one can consider the combination of $\mup\neum$ as the particle $X$ and treat \Dsm as $A$. Although \Dsm is long-lived, the decay vertex of the \Bs meson is constrained by the \Dsm and \mup, thus our technique is still applicable. 

\begin{figure}
  \begin{center}
  \includegraphics[width=0.5\textwidth]{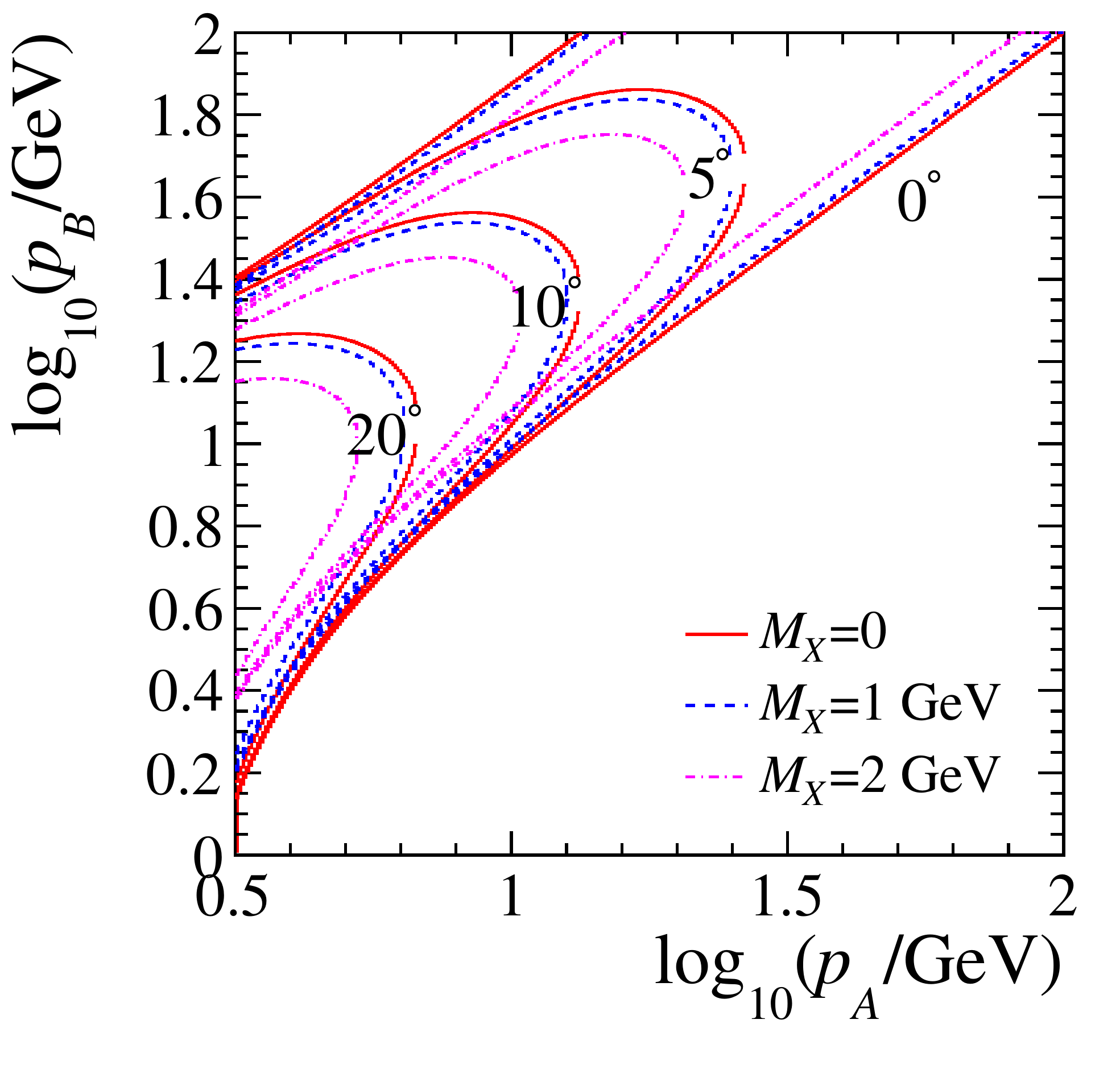}
  \end{center}
  \caption{Logarithms of the magnitudes of the reconstructed momentum $p_B$ of the \Bs meson as functions of the momentum $p_A$ of the 
   reconstructed state $A$ of mass 2\gev for the different angles $\theta$ and masses $m_{X}$ of the invisible state $X$. The $p_B$ solutions for each value of $p_A$ are plotted following 
   Eq.~(\ref{eq:bs_mom}). 
  }
  \label{fig:bs_mom}
\end{figure}

Under the conditions formulated above, the momentum of the $\Bs$ meson can be reconstructed from the momentum of the system $A$ ($p_A$) and 
the angle between the directions of $A$ and $\Bs$ in the laboratory frame ($\theta$) up to a two-fold ambiguity: 
\begin{equation}
  p_{B} = \frac{\left(M_{B}^{2} + \Delta \right)p_{A}\cos\theta \pm E_A
  \sqrt{\Delta^2 + M_{B}^2(M_{B}^{2} + \Delta - 4 M_{A}^{2} - 4 p_{A}^{2}\sin^2\theta)}
  }{2(M_{A}^{2} + p_{A}^{2}\sin^2\theta)}, 
  \label{eq:bs_mom}
\end{equation}
where $M_{B}$, $M_{A}$ and $M_{X}$ are masses of the $\Bs$, $A$ and $X$ states, respectively, and $\Delta$ and $E_A$ are defined as 
\begin{equation*}
  \begin{split}
    \Delta & \equiv M_A^2 - M_X^2, \\
    E_A & \equiv \sqrt{M^2_A + p^2_A}. 
  \end{split}
\end{equation*}
Figure~\ref{fig:bs_mom} demonstrates the behaviour of $p_B$ as a function of $p_A$ and the angle $\theta$
for fixed $M_A = 2\gev$. 
The two solutions for the momentum are denoted in the following text as $p^{\rm (min)}_B$ and $p^{\rm (max)}_B$. 

The observation of \Bs oscillations requires the knowledge of the $\Bs$ momentum $p_B$, as well as the 
decay length $L$. This gives the decay time
\begin{equation}
  t = L \frac{M_{B}}{p_{B}}.
  \label{eq:decay_time}
\end{equation}
Oscillations result in a modulation of the rate of flavour-tagged \Bs decays. 
Assuming perfect flavour tagging, the decay time distributions for the mixed and unmixed \Bs decays are
\begin{equation}
  f_{\pm}(t) \propto e^{-\Gamma_s t}\left[ \cosh\left(\frac{\Delta\Gamma_s t}{2}\right) \pm \cos(\Delta m_s t)\right], 
  \label{eq:tagged_decay_time}
\end{equation}
where the sign ``$+$'' corresponds to unmixed and ``$-$'' to mixed decays. 
Here, $\Delta\Gamma_s$ is the width difference between the heavy and the light \Bs mass eigenstates, 
and $\Delta m_s$ is their mass difference, which drives the oscillations. 

In the case of partially reconstructed \Bs decays, one can use the topologically reconstructed momentum $p_B$ from Eq. (\ref{eq:bs_mom})
in the calculation of decay time. Since there are two solutions for $p_B$, a sample of partially reconstructed decays will provide
two decay time distributions, $f^{\rm (min)}(t)$ and $f^{\rm (max)}(t)$. Only one of the solutions corresponds to the true \Bs momentum for each decay, 
thus wrong solutions will not produce a coherent oscillation pattern and will result in a smooth background contribution in 
both densities. 

Once the correct mass $M_X$ is taken in the calculation of $p_B$, the distribution of the reconstructed decay 
time will contain an oscillating term with the frequency corresponding to the mass difference $\Delta m_s$. In what follows, lowercase $m_X$ is used to distinguish the $X$ mass assigned in the calculation of $p_B$ from its true mass $M_X$. 
A biased value of mass $m_X\neq M_X$ will result in the oscillating term having higher or lower frequency. One can scan the values of the assigned mass $m_X$ and calculate the amplitude of the harmonic corresponding to the mass difference $\Delta m_s$ in the spectrum of reconstructed decay times. The maximum of the amplitude will correspond to the true mass $m_X = M_X$.

Instead of the absolute rates of the $\Bs$ and $\Bsb$ mesons, it is convenient to study the flavour asymmetry
\begin{equation}
  a_{\Bs}(t) = \frac{N_{\rm unmix}(t) - N_{\rm mix}(t)}{N_{\rm unmix}(t) + N_{\rm mix}(t)}, 
  \label{eq:fl_asym}
\end{equation}
where $N_{\rm mix}$ and $N_{\rm unmix}$ are the mixed and unmixed yields as a function of decay time. The asymmetry 
oscillates as a function of decay time. If the correct momentum is used to calculate $t$ from the decay length $L$, then the oscillations of the flavour asymmetry will occur with known frequency $\Delta m_s$. 

The complex amplitude $A_{\Bs}$ of the harmonic corresponding to the measured $\Bs$ oscillation frequency,
\begin{equation}
  A_{\Bs} \equiv C_{\Bs} + iS_{\Bs} = \int a_{\Bs}(t) e^{i\Delta m_s t}\;dt, 
  \label{eq:bs_harmonic}
\end{equation}
with the real $C_{\Bs}$ and imaginary $S_{\Bs}$ parts, will therefore probe the 
correctness of the mass hypothesis for the $X$ state. Here $A_{\Bs}$ depends on $m_X$ indirectly, through the dependence of reconstructed decay time $t$ on the momentum $p_B$
which, in turn, depends on $m_X$. 
For the correct mass $m_X = M_X$, $C_{\Bs}(m_X)$ should reach maximum, while $S_{\Bs}(m_X)$ should cross 
zero\footnote{Assuming $CP$ conservation in the interference of the $\Bs$ mixing and $\Bs\to AX$ decay. }. 

An attractive feature of this approach is that the measurement of the amplitude of high-frequency 
oscillations does not require detailed knowledge of the low-frequency backgrounds coming from 
non-\Bs decays or random combinations of particles produced in the primary vertex of a proton-proton interaction. As a result, this method is expected to have low 
systematic uncertainty due to the background description. 

If several invisible states $X_i$ contribute to the \Bs decay, the mass spectrum $A_{\Bs}(m_X)$ will 
consist of several peaks at the masses of each $X_i$ state. The relative magnitudes of the peaks will give relative 
branching ratios to the final states with $X_i$ states. Finally, if the mass spectrum of $X$ is continuous, 
the shape of the function $A_{\Bs}(m_X)$ will be the convolution of the single-mass response $A_{\Bs}(m_X|M_X)$ for a given $M_X$ with the invariant mass spectrum of the $X$ system $P(M_X)$. 

Even assuming perfect tagging and vertex resolution, the mass resolution will be limited since it is 
determined by a finite number of observed oscillation periods. However, with a large number of \Bs decays it should be possible to perform a deconvolution of the $A_{\Bs}(m_X)$ function to reconstruct the true $X$ mass spectrum. 
Finally, it is possible that a maximum likelihood fit to the two distributions $f^{\rm (min)}(t)$ and $f^{\rm (max)}(t)$
will provide more information than the study of the Fourier spectrum of \Bs flavour asymmetry. This, however, would require 
a good control of slowly varying components of the decay time distributions. 

While the observation of oscillations in the flavour asymmetry provides a distinctive experimental signature only for the fast \Bs oscillations, the relation between decay time and mass spectrum given by Eqs.~(\ref{eq:bs_mom}) and (\ref{eq:decay_time}) can provide information for partially reconstructed decays of other weakly decaying hadrons. Such a generalisation would require a completely different study, possibly using real data in the application to a specific final state, since it will be strongly affected by various backgrounds
and acceptance effects, and is therefore beyond the scope of this paper. 

\section{Simulation studies}

\label{sec:simulation}

Simplified Monte Carlo simulations are performed to test the feasibility of the proposed approach. The kinematic properties of two-body $\Bs\to AX$
decays are simulated with $M_A=2\gev$ and $M_X$ varying between 0 and $2\gev$. The initial \Bs mesons are generated with momentum and pseudorapidity distributions roughly matching those obtained in the full and simplified LHCb simulation~\cite{Cowan:2016tnm} for the decay modes containing only hadrons in the final state. The transverse momentum \pt is generated following an exponential distribution with a slope of $5\gev$ over the threshold of 8\gev. The distribution of $\eta$ is a Gaussian function with the mean of 3.2 and the standard deviation of 0.46. The decay time distribution follows Eq.~(\ref{eq:tagged_decay_time}). All the following studies assume perfect flavour tagging; the effect of realistic flavour tagging performance can be taken into account by scaling the size of the data sample. 

\begin{figure}
  \includegraphics[width=\textwidth]{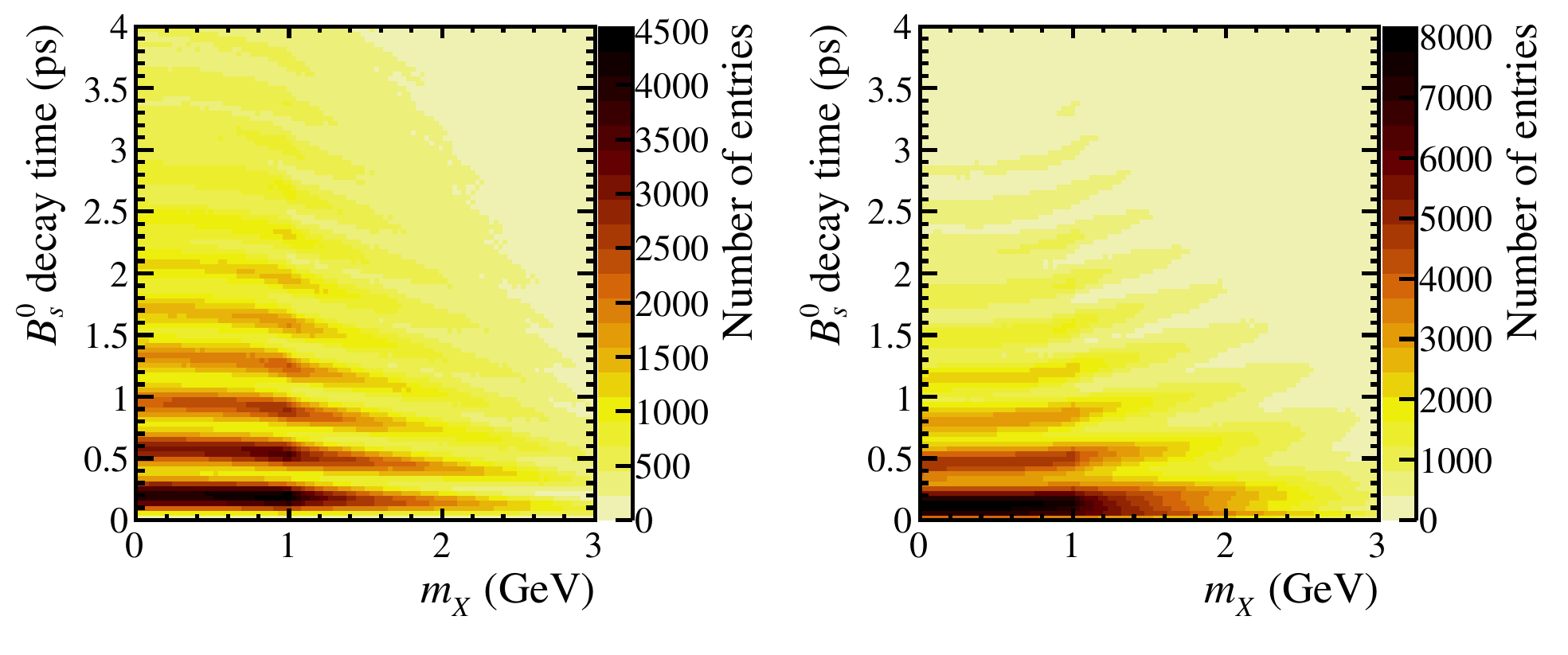}
  \put(-310,160){\colorbox{white}{(a)}}
  \put(-83,160){\colorbox{white}{(b)}}
  \caption{$\Bs$ decay time reconstructed from decay length using $p^{\rm (min)}_{B}$ (a) and $p^{\rm (max)}_{B}$ (b)
           solutions as a function of 
           the mass $m_X$ used in topological reconstruction. 
           The true mass of the $X$ state is $M_X=1\gev$. }
  \label{fig:osc_plot}
\end{figure}

For each of the generated \Bs decays, the momentum (and subsequently the decay time) is calculated from topological information following Eqs.~(\ref{eq:bs_mom}) and (\ref{eq:decay_time}). Figure~\ref{fig:osc_plot} shows the distributions of \Bs decay time as a function of mass $m_X$ 
for the decays generated with $M_X=1\gev$. Each plot corresponds to a solution of Eq.~(\ref{eq:bs_mom}) for the \Bs momentum. 
The dependence of the oscillation frequency on the assigned mass is clearly seen. 
Once the distributions of the decay time calculated this way for a certain 
$m_X$ are obtained for the mixed and unmixed decays, the flavour asymmetry as a function of decay time is calculated according to Eq.~(\ref{eq:fl_asym}). Only the solution $p_B^{\rm (min)}$ is used since it more often gives the right answer for higher \Bs momenta (and thus higher decay times). An approach using multivariate regression based on 
event topology information~\cite{Ciezarek:2016lqu} can be used to make the choice of the correct solution more efficient. 

\begin{figure}
  \centering
  \includegraphics[width=0.5\textwidth]{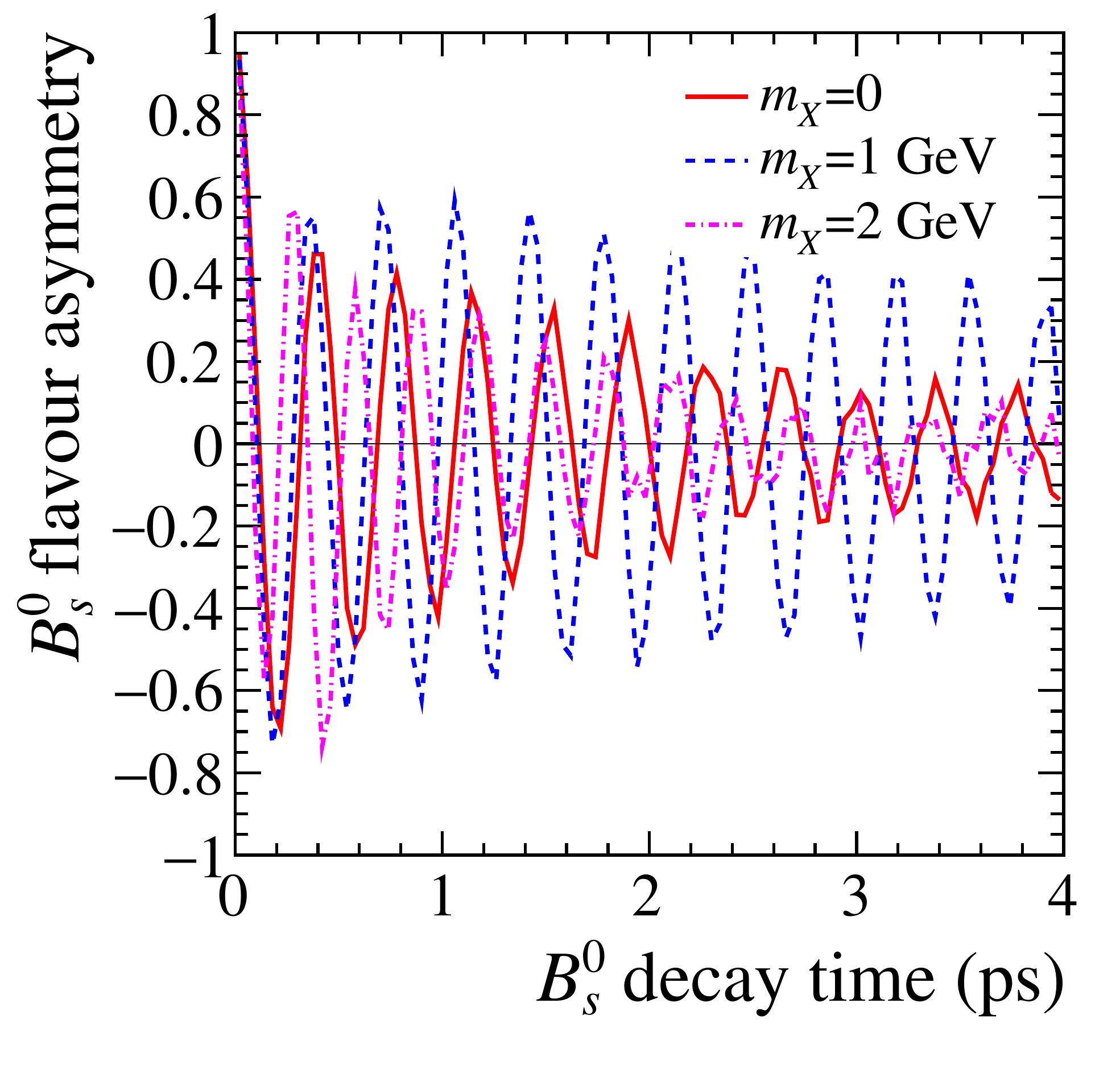}
  \caption{Dependence of $\Bs$ flavour asymmetry on decay time $t$ for the true $X$  mass $M_X=1\gev$, 
           for different assigned masses $m_X$.} 
  \label{fig:asym_plot} 
\end{figure}

Figure~\ref{fig:asym_plot} shows the flavour asymmetry as a function of decay time for a signal-only sample of $10^5$ $\Bs\to AX$ events of each initial \Bs flavour
with true $M_X = 1\gev$ and taking different assigned masses $m_X$ of 0, 1, and 2\gev. Perfect vertex resolution is assumed when 
calculating the decay time. The amplitude of flavour asymmetry oscillations is maximal for the correct mass hypothesis, while for 
the wrong masses, the coherence at higher decay times gets worse. In addition, the correct mass hypothesis yields a frequency of oscillations
corresponding to the true mass difference $\Delta m_s$, while the wrong mass hypotheses yield higher or lower frequencies. 

\begin{figure}
  \centering
  \includegraphics[width=0.43\textwidth]{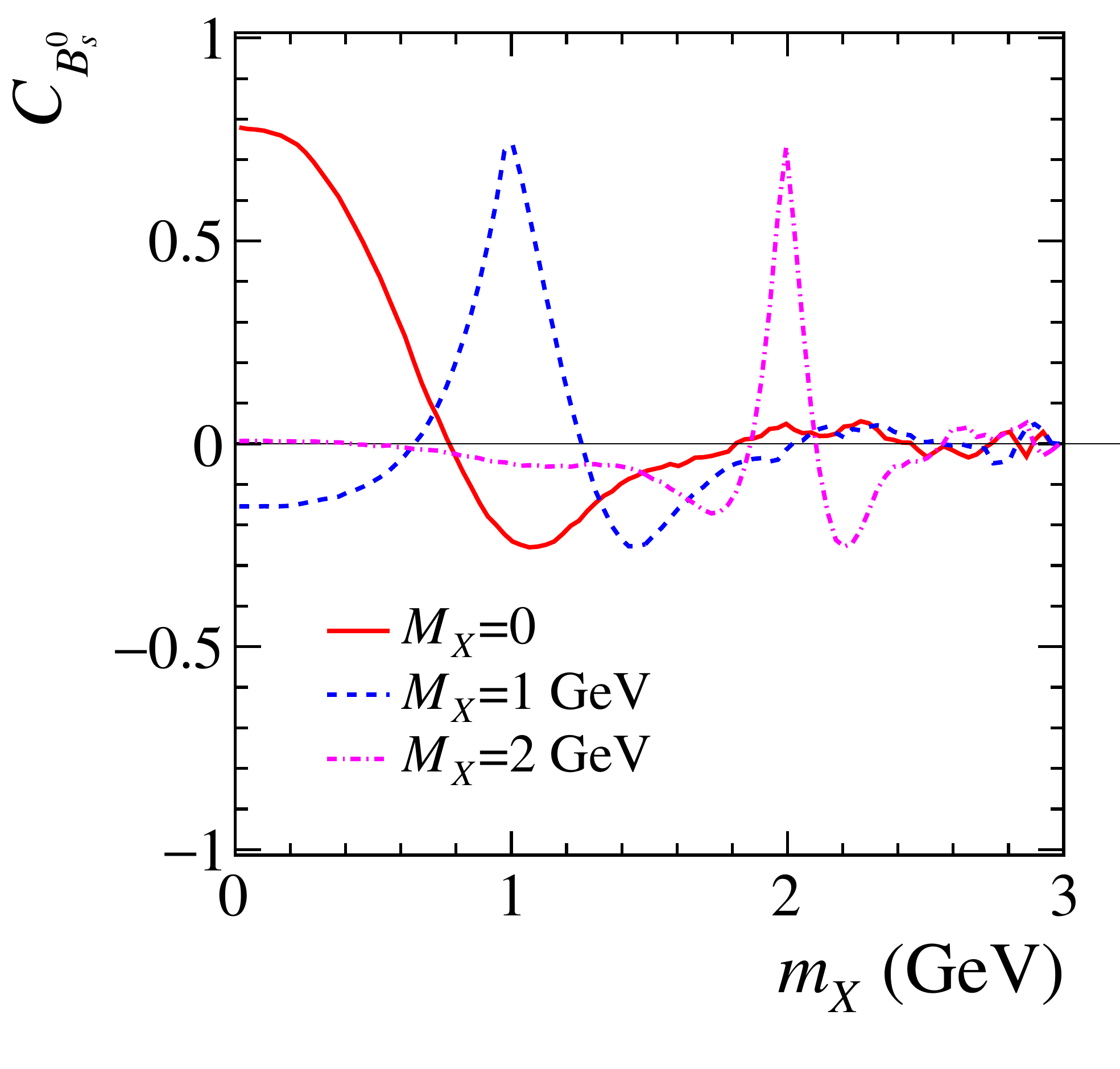}
  \put(-35, 48){(a)}
  \includegraphics[width=0.43\textwidth]{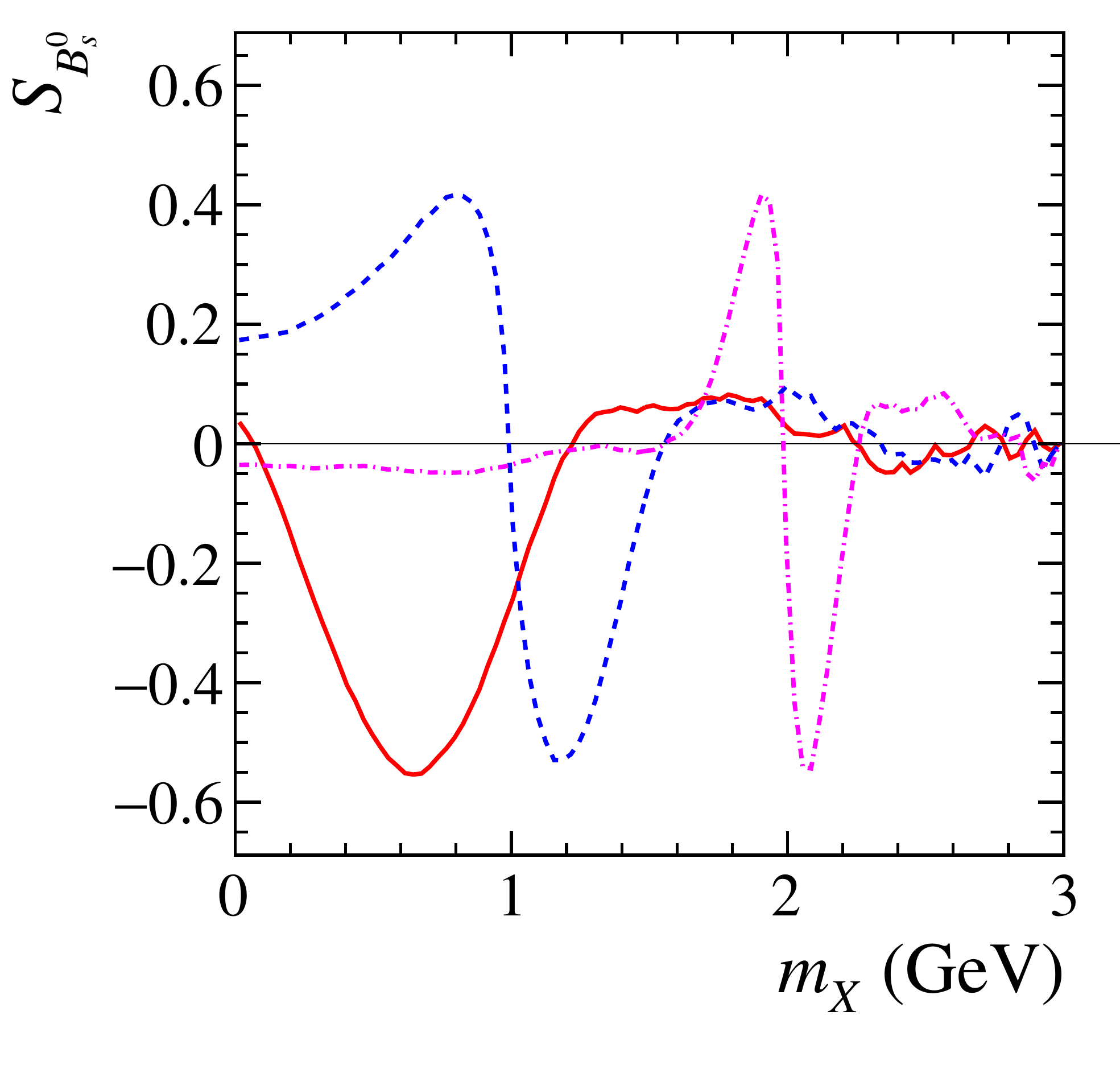}
  \put(-35, 48){(b)}
  \caption{(a) Real $C_{\Bs}$ and (b) imaginary $S_{\Bs}$ parts of the amplitude of the harmonic corresponding to \Bs oscillation frequency 
           $\Delta m_s$ as a function of assigned mass $m_X$ for three values of the true mass $M_X$, for perfect $\Bs$ decay vertex 
           resolution and $10^{5}$ events of each \Bs flavour. }
  \label{fig:mass_spec_nores}
\end{figure}

One then obtains the spectrum of $X$ masses by scanning $m_X$ in the range between 0 and 3\gev and calculating the amplitude 
of the \Bs oscillation harmonic (Eq.~(\ref{eq:bs_harmonic})) in the flavour asymmetry distribution. 
The selection of \Bs mesons in a typical analysis involves the requirement of non-zero lifetime and the reconstruction efficiency at low decay times is small (see Fig.~\ref{fig:bs_mixing}). Thus, only the decays with $t>0.8\ps$, corresponding to around half of of the \Bs lifetime, are used to calculate the amplitude of the \Bs oscillation harmonic. 
The $X$ mass spectra (in the form of real and imaginary amplitudes as a function of $m_X$) for true masses 
$M_X$ of 0, 1, and 2\gev are shown in Fig.~\ref{fig:mass_spec_nores}. The mass resolution is better for higher-mass states; this is expected since $M_X$ variations at higher masses lead to greater variations of momentum $p_B$ as follows from Eq.~(\ref{eq:bs_mom}) and is illustrated by Fig.~\ref{fig:bs_mom}. The mass resolution is found to improve for higher $M_A$ masses for the same reason. 
While the spectra presented in the plot contain uncertainties, these are not straightforward to take into account: each 
value of the amplitude is calculated from the same data set, so all the values of the amplitude are correlated. 
The uncertainties are therefore not plotted in Fig.~\ref{fig:mass_spec_nores}. The treatment of experimental uncertainties is discussed in detail in Section~\ref{sec:practical}. 

\begin{figure}
    \centering
    \includegraphics[width=0.43\textwidth]{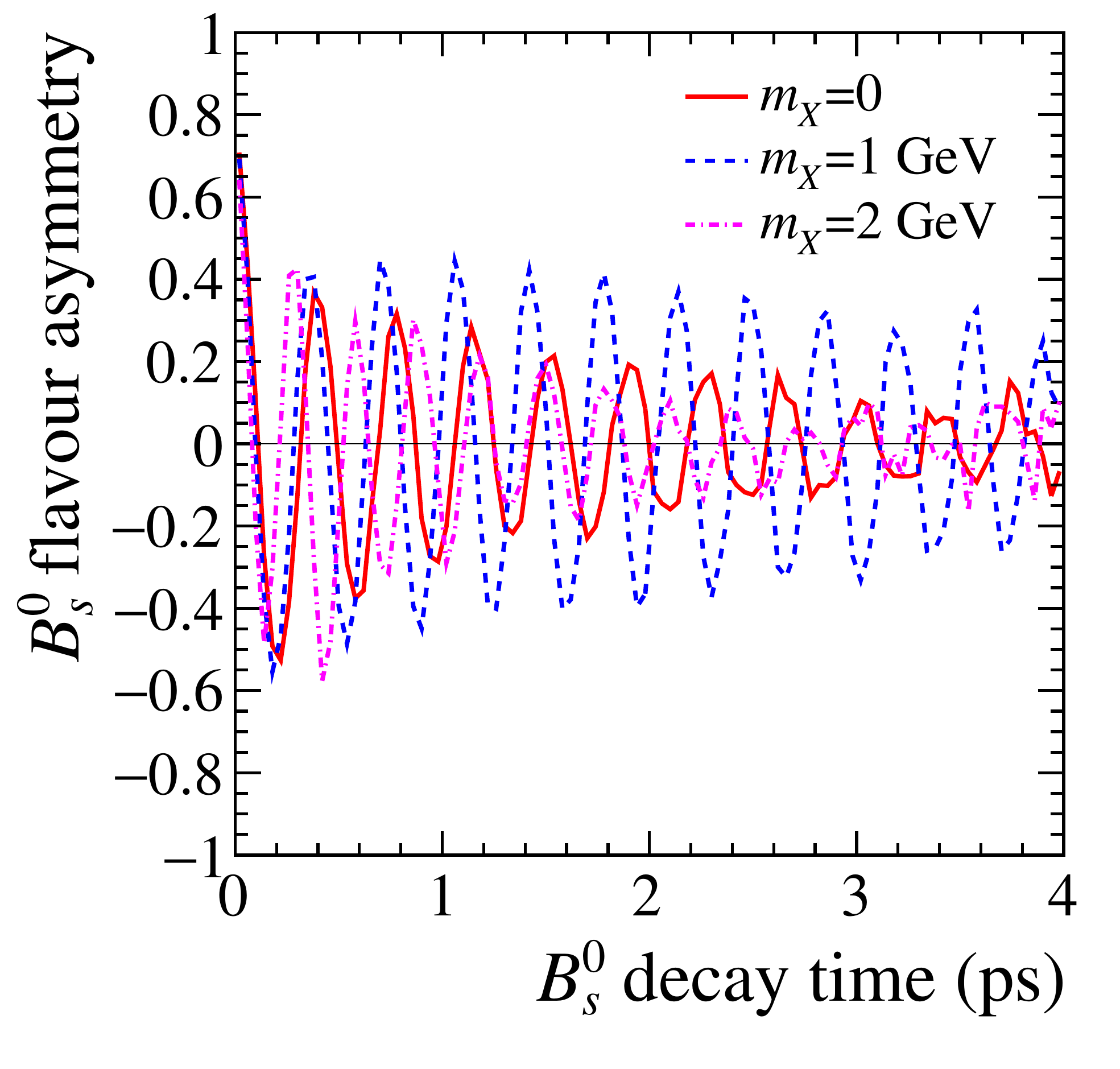}
    \put(-35, 48){(a)}
    \includegraphics[width=0.43\textwidth]{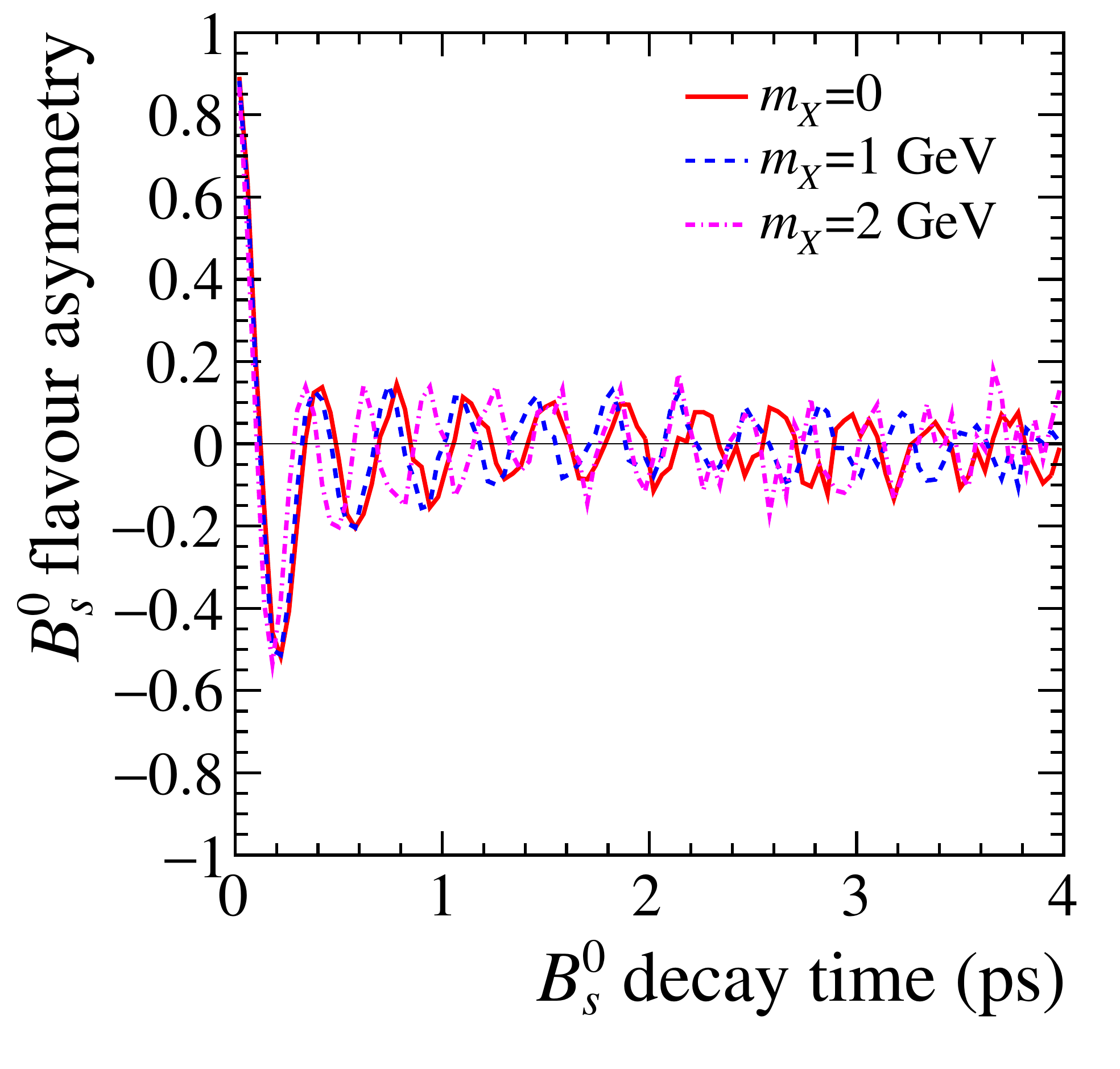}
    \put(-35, 48){(b)}
    \caption{Dependence of $\Bs$ flavour asymmetry on decay time $t$ for the true $X$ mass $M_X=1\gev$, for different assigned masses $m_X$, with (a) non-zero longitudinal vertex resolution $\sigma_z=300\mum$ and perfect transverse resolution, and (b) non-zero transverse vertex resolution $\sigma_x=40\mum$ and perfect longitudinal resolution. }
    \label{fig:osc_asym_res}
\end{figure}

Good vertex resolution is an important factor for the performance of the proposed technique. Finite precision in the reconstruction of the position of the \Bs decay vertex transverse to its flight direction (which is approximately equal to the component $\sigma_x$ of the vertex resolution perpendicular to the direction of proton beams, since $\Bs$ mesons are boosted along the beams) translates to the uncertainty of the angle $\theta$ and thus of the momentum $p_B$. The uncertainty in $p_B$ results in a loss of coherence at higher 
decay times, which in turn leads to a worse $X$ mass resolution. 
Finite vertex resolution in the direction along the beams (longitudinal resolution) $\sigma_z$ affects the measurements to a lesser extent: once the resolution is small compared to the oscillation period, it results in a certain reduction of the flavour asymmetry amplitude. The vertex resolution at LHCb depends on the charge multiplicity and kinematics of the decay, with typical values for a transverse component $\sigma_x$ of a few tens of~\mum. The longitudinal component of the vertex coordinate is typically determined with the precision of 
200--300\mum~\cite{LHCbVELOGroup:2014uea}. 
The effect of finite vertex resolution on the flavour asymmetry oscillations is illustrated in Figs.~\ref{fig:osc_asym_res}, separately for the longitudinal component of 300\mum and the transverse component of 40\mum. 

\begin{figure}
  \centering
  \includegraphics[width=0.41\textwidth]{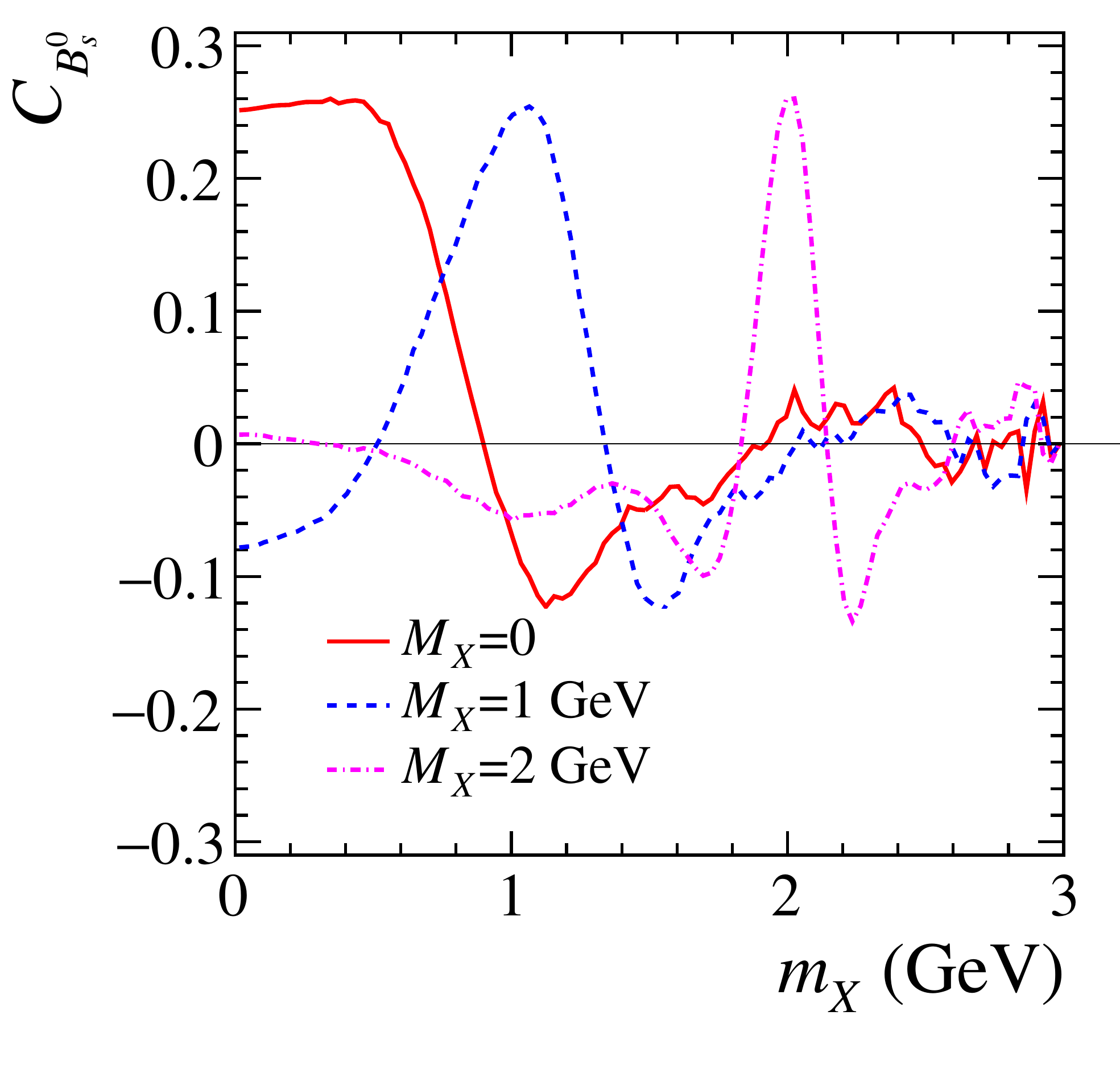}
  \put(-35, 48){(a)}
  \includegraphics[width=0.41\textwidth]{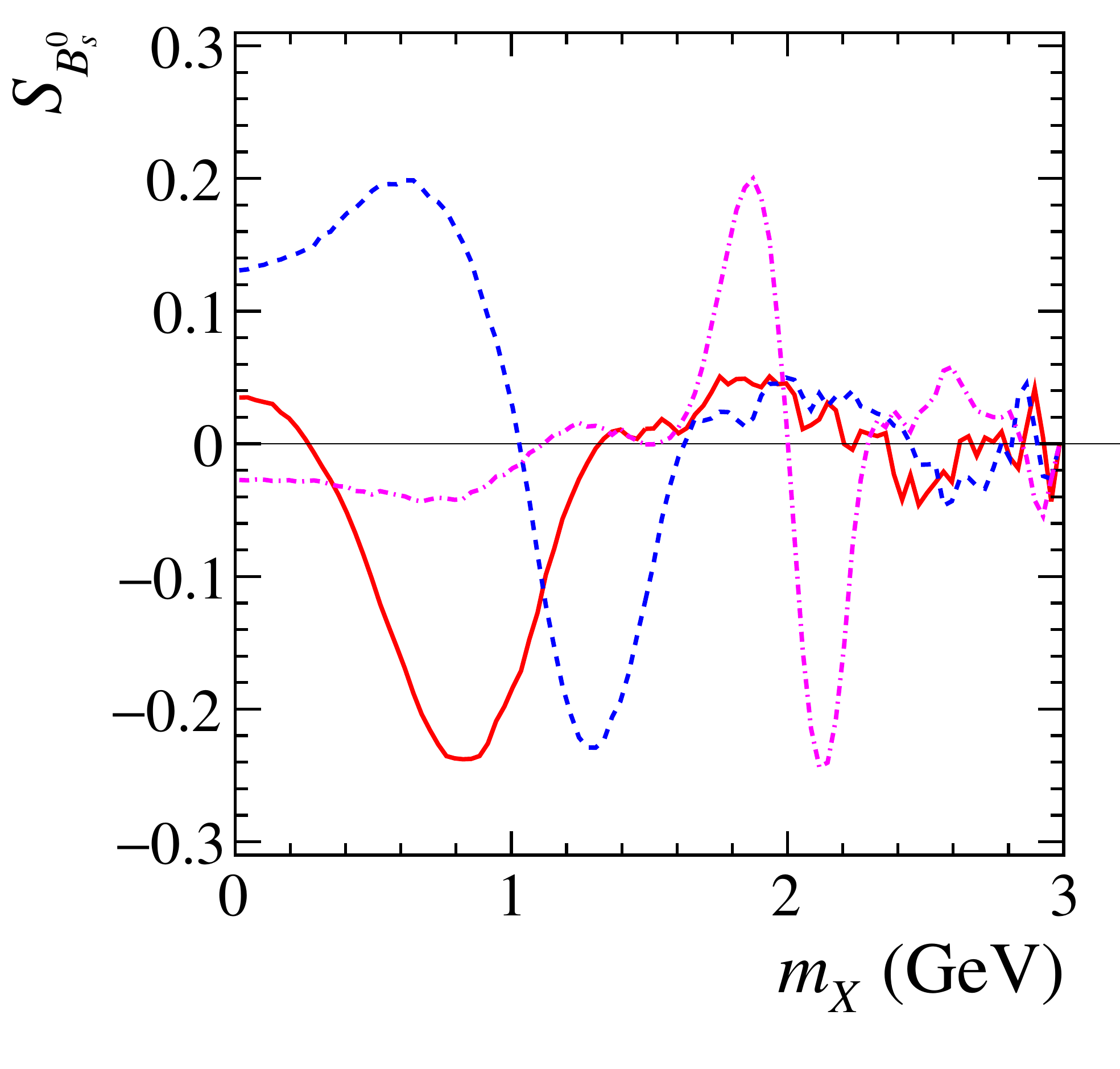}
  \put(-35, 48){(b)}

  \includegraphics[width=0.41\textwidth]{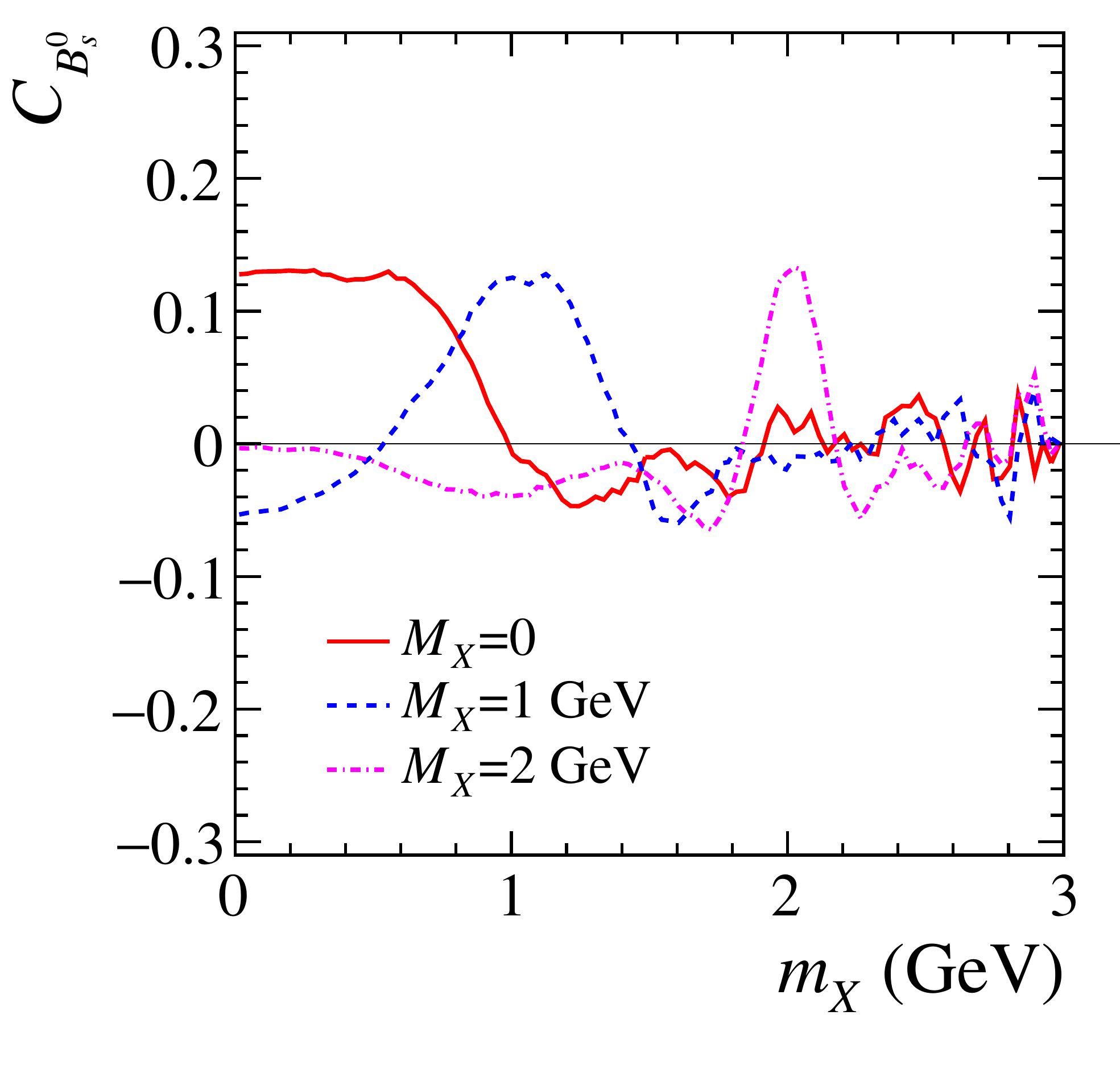}
  \put(-35, 48){(c)}
  \includegraphics[width=0.41\textwidth]{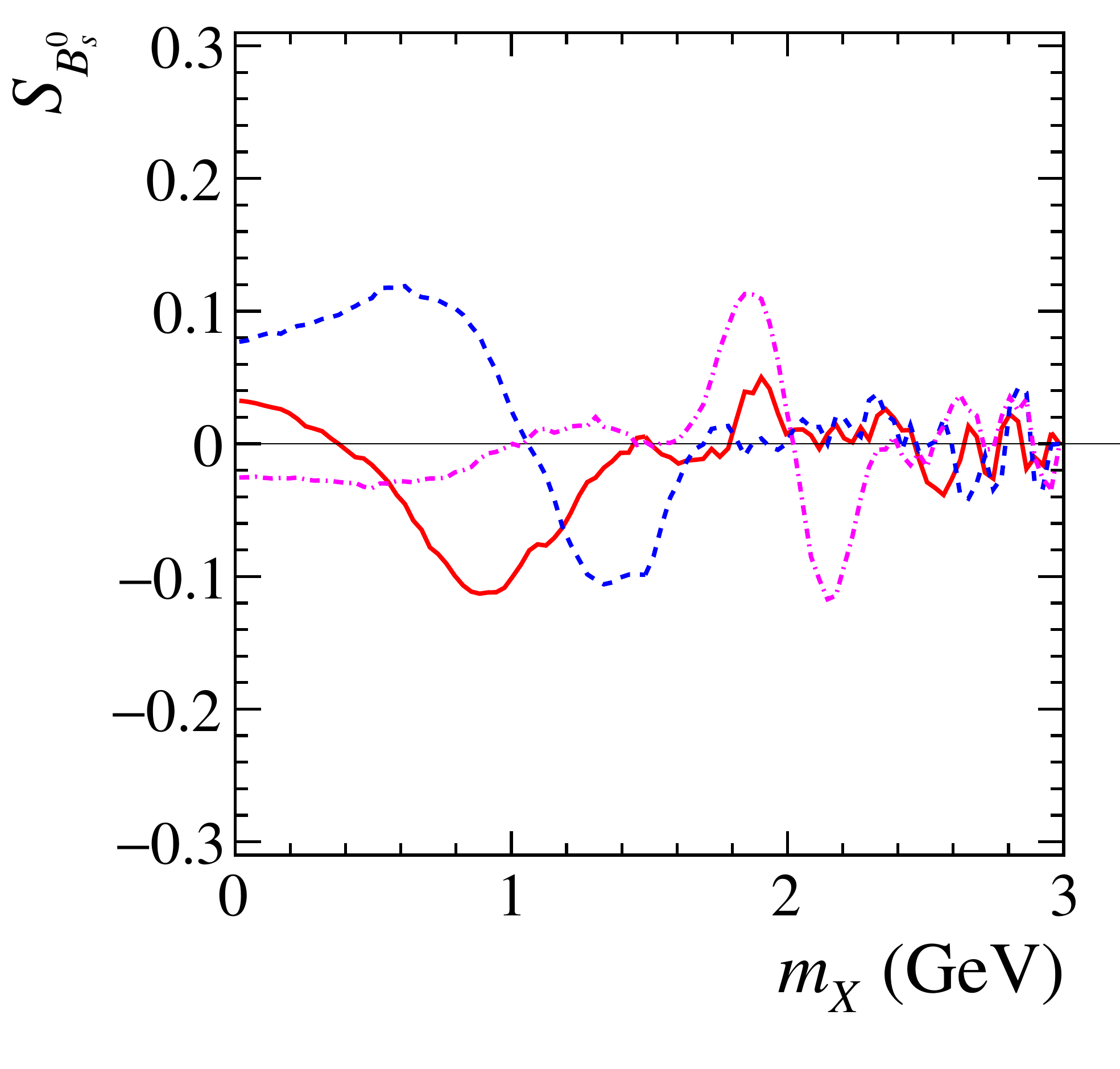}
  \put(-35, 48){(d)}

  \includegraphics[width=0.41\textwidth]{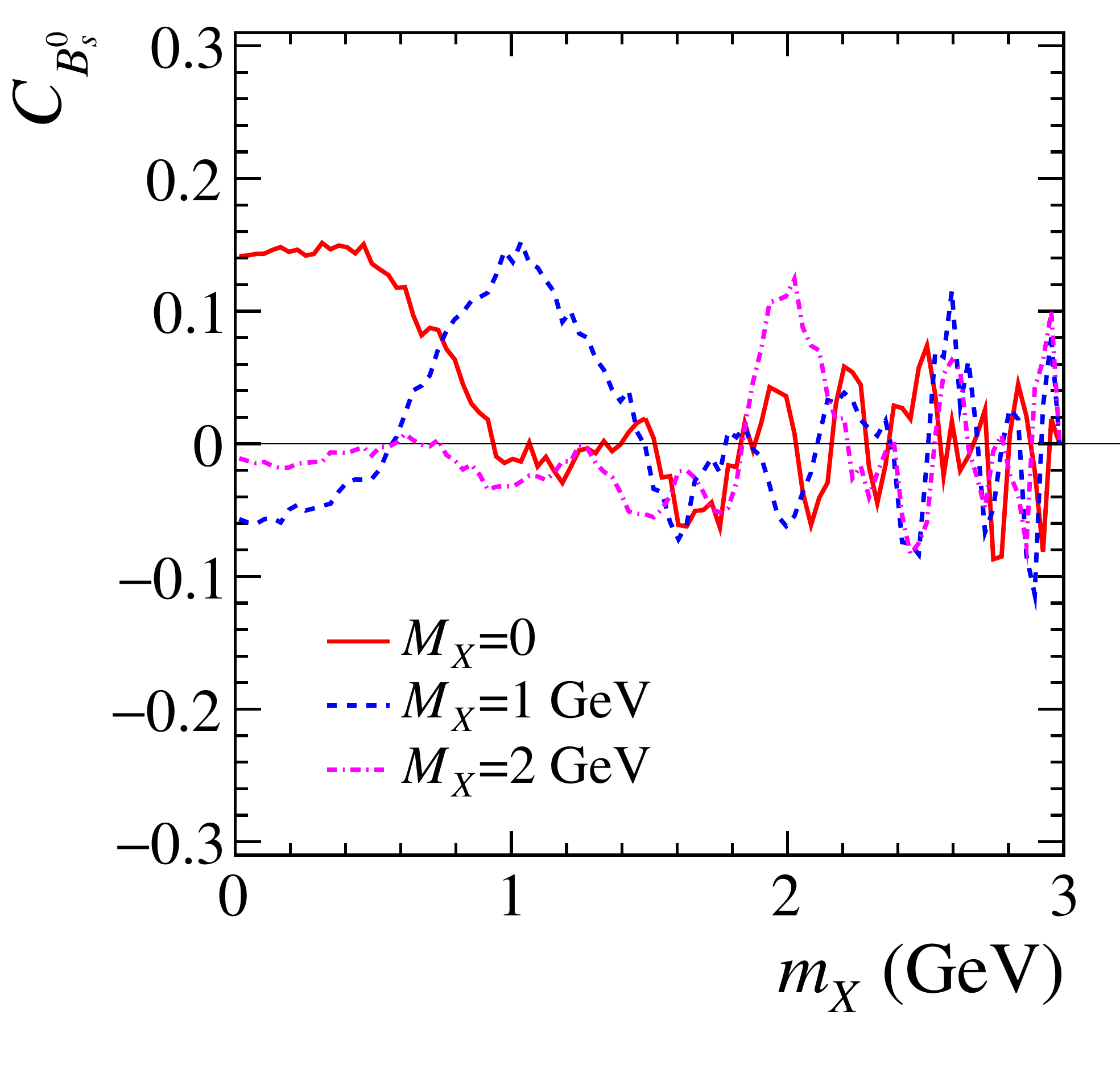}
  \put(-35, 48){(e)}
  \includegraphics[width=0.41\textwidth]{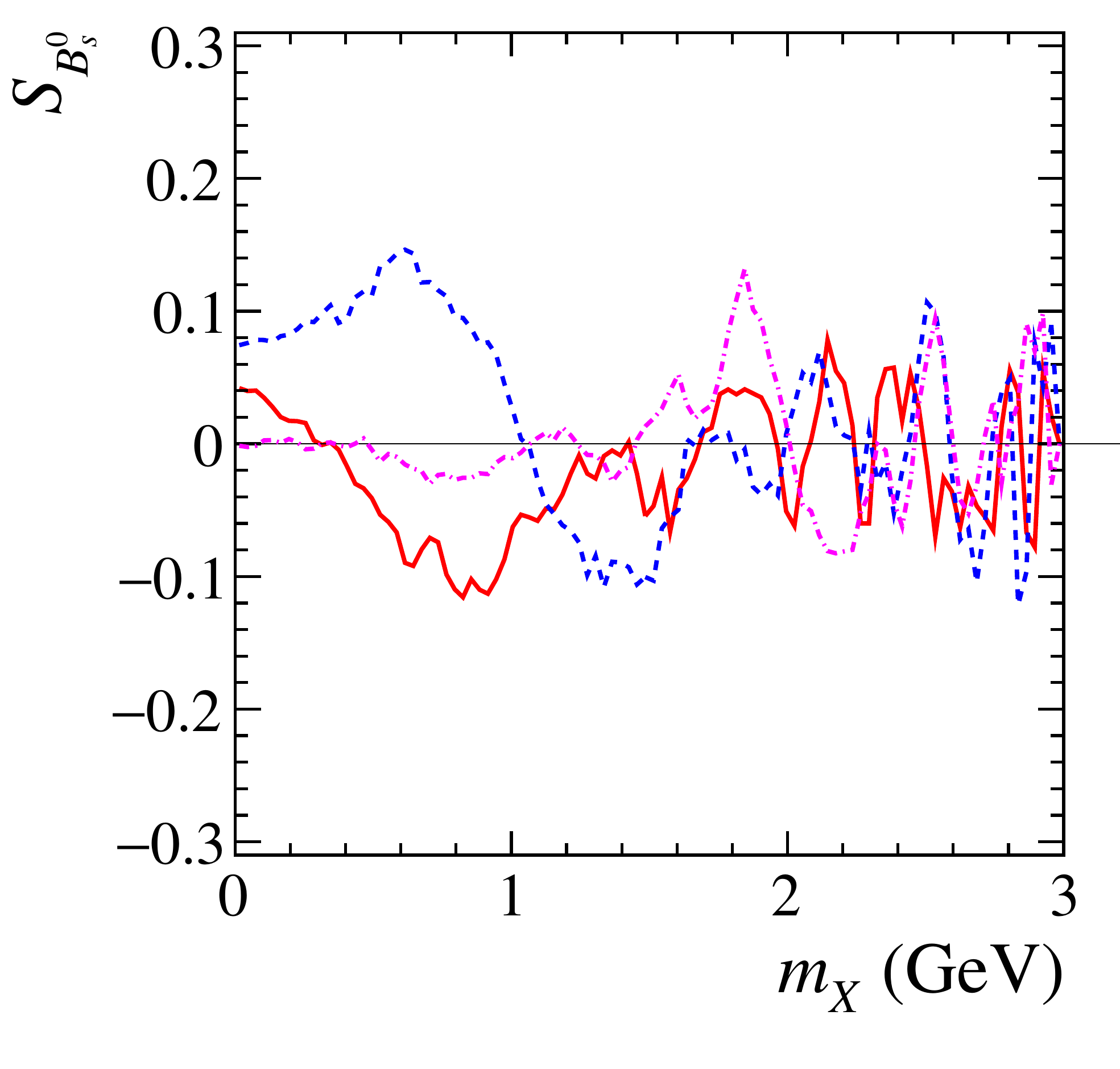}
  \put(-35, 48){(f)}
  \caption{(a,c,e) Real $C_{\Bs}$ and (b,d,f) imaginary $S_{\Bs}$ parts of the amplitude of the harmonic 
           corresponding to \Bs oscillation frequency $\Delta m_s$ as a function of 
           assigned mass $m_X$ for three values of the true mass $M_X$, 
           for (a,b) vertex resolution components $\sigma_x = 20\mum$, $\sigma_z = 0$ and $10^{5}$ events of each \Bs flavour, 
               (c,d) $\sigma_x = 40\mum$, $\sigma_z = 0$ and $10^{5}$ events of each \Bs flavour, 
           and (e,f) $\sigma_x = 40\mum$, $\sigma_z=300\mum$ and $10^{4}$ events of each \Bs flavour. }
  \label{fig:mass_spec_res}
\end{figure}

Figures~\ref{fig:mass_spec_res}(a--d) demonstrate the effect of a gradual increase of the transverse resolution to 20\mum and 40\mum on the mass spectra $C_{\Bs}$ and $S_{\Bs}$. These plots use the same sample size of $10^5$ as for Fig.~\ref{fig:mass_spec_nores}. 
Figures~\ref{fig:mass_spec_res}(e,f) show the mass spectra obtained with vertex resolution $\sigma_x=40\mum$ and $\sigma_z=300\mum$, and using the smaller sample of $10^4$ perfectly tagged \Bs decays of each flavour. Taking the typical effective flavour tagging efficiency at LHCb of 6\%~\cite{Aaij:2017lff} and assuming the statistical uncertainties to scale as inverse square root of the effective tagging efficiency, this corresponds to around $8\times 10^{4}$ reconstructed \Bs decays of both flavours. This is a realistic number of decays with, \eg, single-charm final states for the currently available LHCb data set recorded during the 2011--2018 data taking period. The possible benchmark mode  $\Bs\to\Dsm\proton\overline{n}$ mentioned above should have sufficient yield in the currently accumulated LHCb data sample. 

\section{Practical considerations}

\label{sec:practical}

Analyses using the technique proposed above with experimental data will need to deal with such effects as background contamination, imperfect flavour tagging, finite vertex resolution and non-uniform acceptance, as well as be able to provide statistical and systematic uncertainties of the measured parameters. The exact approaches to be used will depend strongly on the decay mode under study. In this section, possible ways to address these issues are demonstrated using the same fictional two-body decay $\Bs\to AX$ with the fully reconstructed state $A$ of mass 2\gev and the unreconstructed state $X$ of a well-defined mass. 

There are several possible background contributions that may affect the correctly tagged $\Bs\to AX$ signal considered in Section~\ref{sec:simulation}. One source of backgrounds is the events where the flavour tag is uncorrelated with the flavour of the fully reconstructed $A$ state (incoherent backgrounds). These could be due either to true signal events with poor tags, or to random combinations of tracks that do not come from a single beauty decay. Since these events will not introduce a flavour asymmetry, their contribution results in a reduction of the statistical precision of the measurement: it can be accounted for as a factor that effectively reduces the signal sample size, similar to the effective tagging efficiency factor introduced to account for imperfect flavour tagging~\cite{Aaij:2017lff}. 

Another source of backgrounds could come from the decays of beauty hadrons other than the $\Bs$. In that case, a non-zero flavour asymmetry may emerge; however, it will either be slowly oscillating as a function of decay time (in the case of $\Bz$ decays) or be constant (for other types of beauty hadrons). This may introduce a systematic bias to the $C_{\Bs}$ and $S_{\Bs}$ terms, which slowly changes as a function of $m_X$, and is clearly distinguishable from the peaking signal of \Bs oscillations. 

The backgrounds from partially reconstructed $\Bs$ decays could constitute a more serious problem for this method. The situation here is similar to the traditional analyses that use invariant masses to statistically separate signal and background components. A dedicated background study would need to be done for every specific final state. The spectra for the well-known decay modes can be obtained from simulation (or even from data if they can be fully reconstructed) and added to the function fitted to the $C_{\Bs}(m_X)$ and $S_{\Bs}(m_X)$ spectra. Depending on the quantum numbers (\eg, baryon or lepton number) and the masses of the unreconstructed states, certain backgrounds might be prohibitively high, but this needs to be studied on a case-by-case basis. 

To demonstrate how one can evaluate the significance of the observation of the $\Bs\to AX$ decay and $X$ mass resolution, and to show the effect of non-$\Bs$ backgrounds, 1000 pseudoexperiments are generated with a sample size of $10^4$, $3\times 10^4$ and $10^5$ decays in each pseudoexperiment. The decays generated are $\Bs\to AX$ with the correct flavour tag for the signal, $\Bs\to AX$ with a random flavour tag for the incoherent background, and $\Bz\to AX$ to illustrate the effect of non-\Bs backgrounds. The mass of the $X$ state is fixed to either $M_X=1\gev$ or $M_X=2\gev$. The vertex resolution is taken to be $\sigma_z=300\mum$ in the longitudinal and $\sigma_x=40\mum$ in the transverse direction. The production properties of the initial \Bs are the same as in Section~\ref{sec:simulation}. 

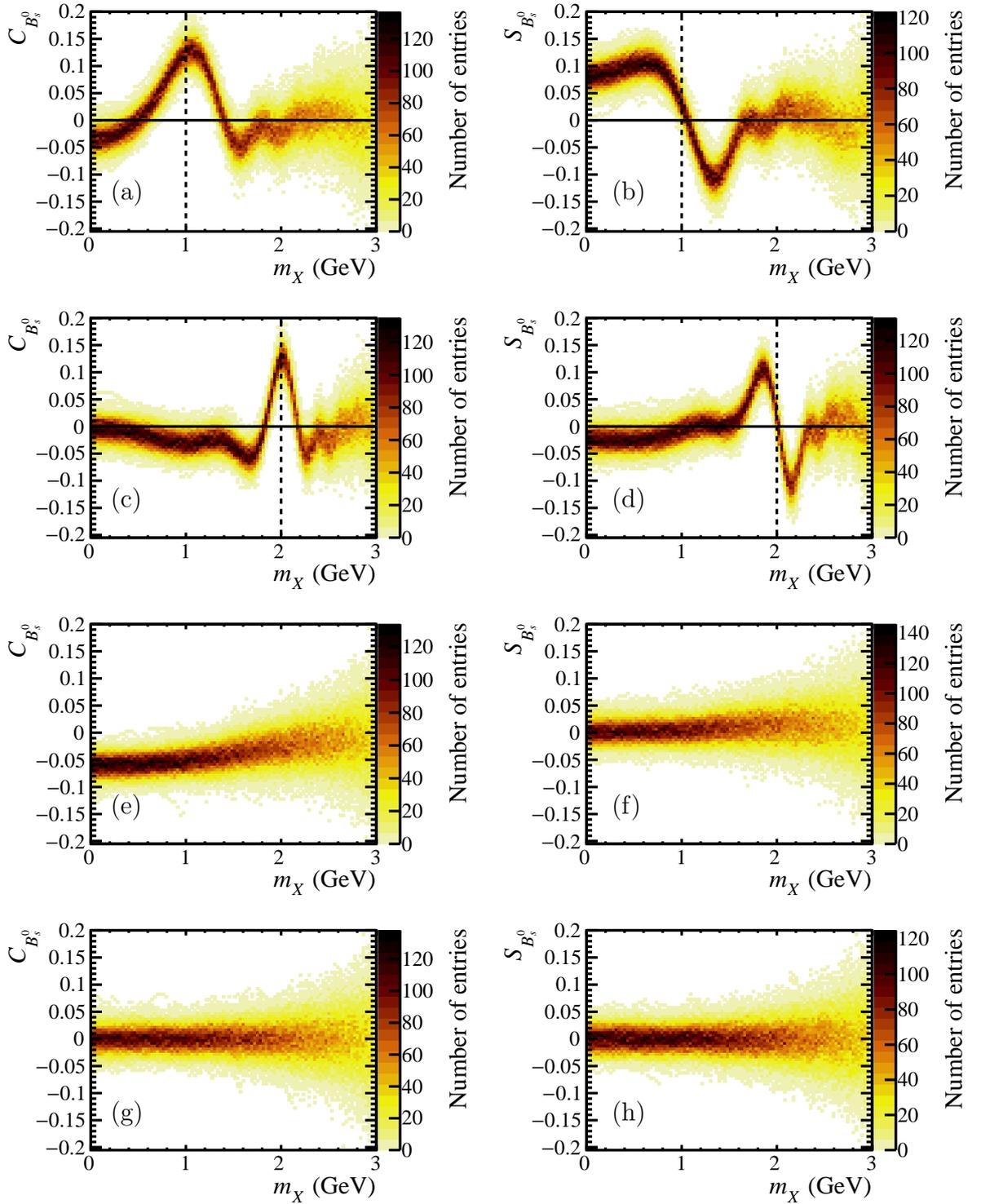
\begin{figure}
    \centering
    \includegraphics[width=\textwidth]{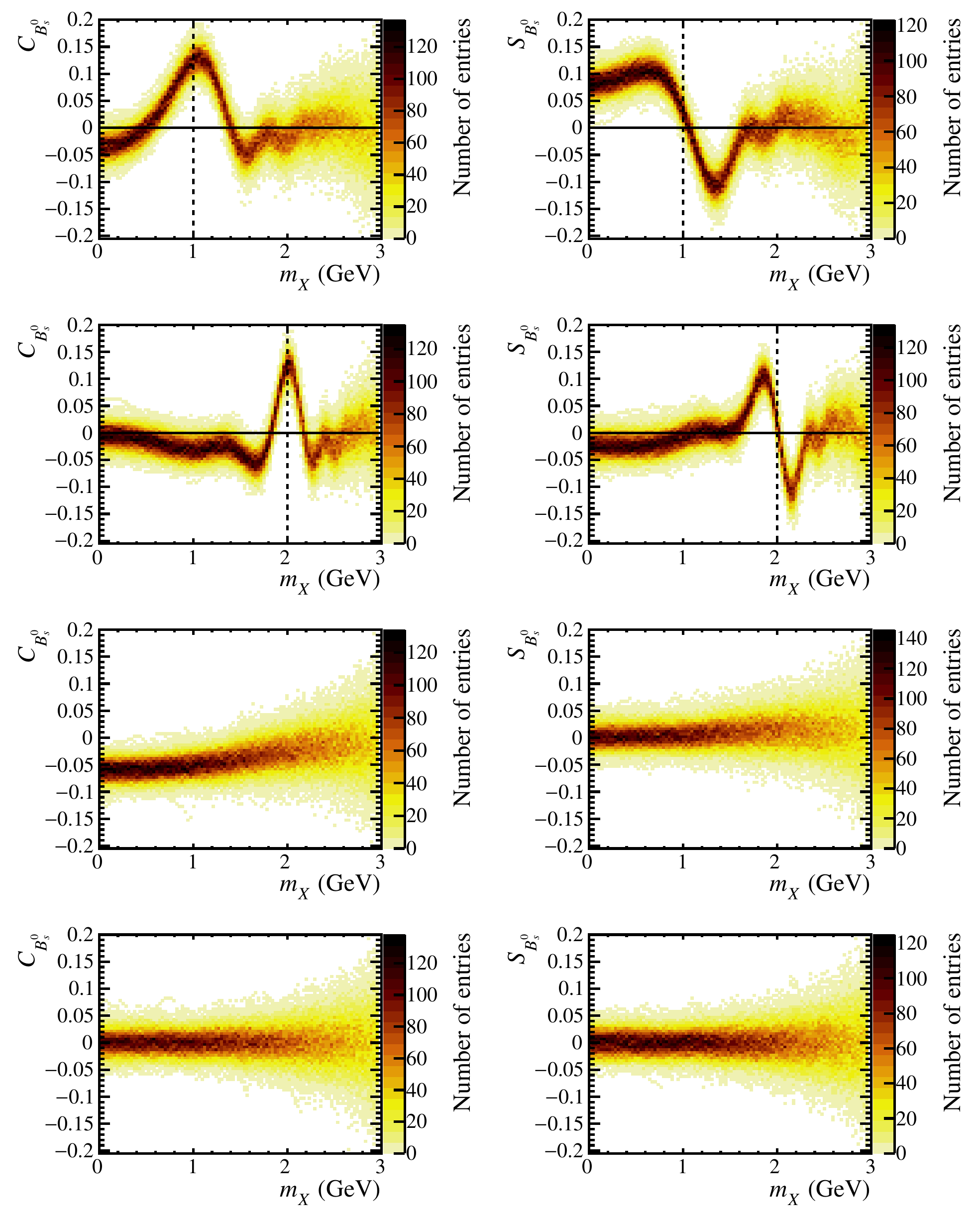}
    \put(-400, 470){(a)}
    \put(-170, 470){(b)}
    \put(-400, 328){(c)}
    \put(-170, 328){(d)}
    \put(-400, 186){(e)}
    \put(-170, 186){(f)}
    \put(-400, 44){(g)}
    \put(-170, 44){(h)}
    \caption{Distributions of the $C_{\Bs}$ and $S_{\Bs}$ terms based on 
    1000 pseudoexperiments of $10^4$ events of each flavour for (a,b)
    $\Bs\to AX$ decay with $M_A=2\gev$ and $M_X=1\gev$, (c,d) the same 
    with $M_X=2\gev$, (e,f) $\Bz\to AX$ decay with $M_X=1\gev$, 
    and (g,h) $\Bs\to AX$ decay with $M_X=1\gev$ and random flavour tag. }
    \label{fig:spec_dist}
\end{figure}

Figure~\ref{fig:spec_dist} shows the distributions of the $C_{\Bs}(m_X)$ and $S_{\Bs}(m_X)$ values collected after adding up all pseudoexperiments with the sample size of $10^4$ events. The spectra are obtained for four cases: $\Bs\to AX$ decay with $M_X=1\gev$~(a,b) and with $M_X=2\gev$~(c,d), for the non-$\Bs$ background decay $\Bz\to AX$~(e,f) and for the incoherent background with the $\Bs$ lifetime, where the flavour of the \Bs is uncorrelated with the final state flavour~(g,h).

\begin{figure}
    \centering
    \includegraphics[width=\textwidth]{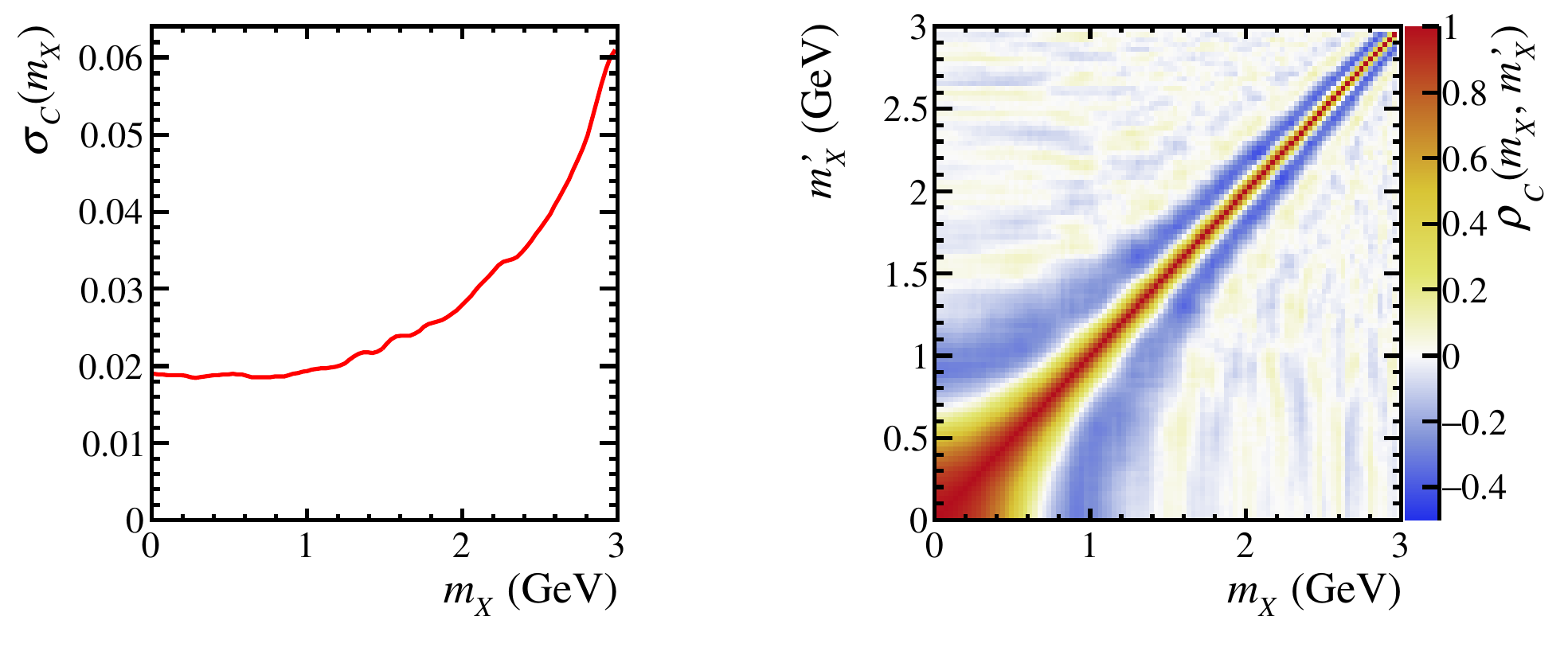}
    \put(-305, 50){(a)}
    \put(-75, 50){\colorbox{white}{(b)}}
    \caption{(a) Statistical uncertainty $\sigma_C(m_X)$ and (b) 
             correlation function $\rho_C(m_X, m'_X)$ of the 
             $C_{\Bs}$ coefficient for the sample of $10^4$ $\Bs\to AX$
             decays of each flavour. }
    \label{fig:sigma_corr}
\end{figure}

As a result, one obtains the distributions of the terms $C_{\Bs}(m_X)$ and $S_{\Bs}(m_X)$ at each point $m_X$. These distributions are well parametrised by a Gaussian function, and the corresponding standard deviation $\sigma_C(m_X)$ as a function of $m_X$ is shown in Fig.~\ref{fig:sigma_corr}(a). The function $\sigma_S(m_X)$ is the same within statistical precision. Moreover, the values of $\sigma_{C,S}(m_X)$ are the same within precision for all four modes considered. The evaluation of the significance of the observed oscillation signal will also need to account for correlations between the measurements of the $C_{\Bs}$ and $S_{\Bs}$ terms at different values of $m_X$. The correlation function $\rho_C(m_X, m'_X)$ obtained from the same pseudoexperiments is shown in Fig.~\ref{fig:sigma_corr}(b); the function $\rho_S(m_X, m'_X)$ is identical within the precision of our simulation. 

\begin{figure}
    \centering
    \includegraphics[width=0.9\textwidth]{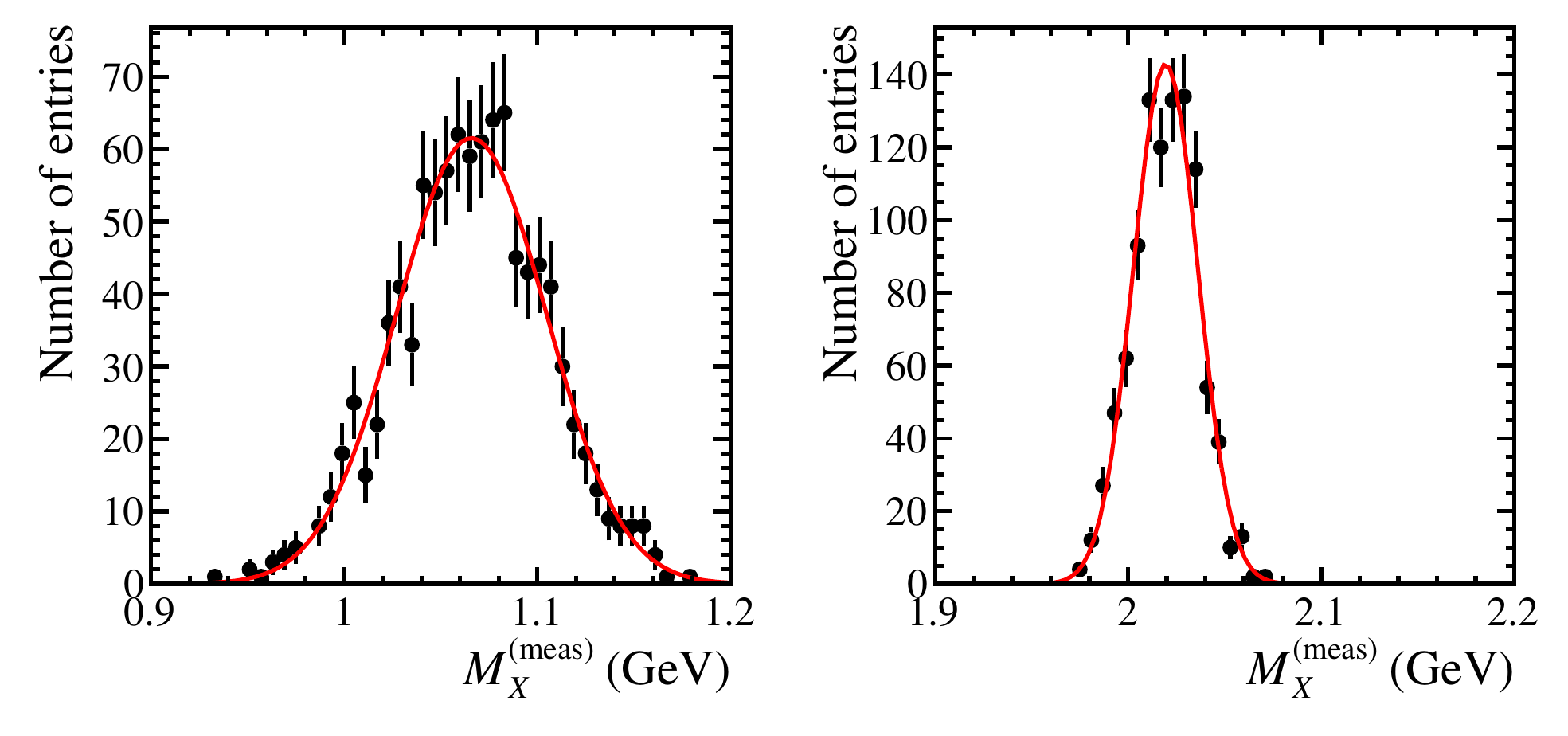}
    \put(-360, 164){(a)}
    \put(-155, 164){(b)}
    \caption{Distributions of the $X$ mass estimate $M_{X}^{(\rm meas)}$, 
             evaluated as the zero crossing point of the $S_{\Bs}(m_X)$
             function for true $X$ mass (a) $M_X=1\gev$ and (b) $M_X=2\gev$. Smooth line is the result of the Gaussian fit. }
    \label{fig:mass_res}
\end{figure}

\begin{figure}
    \centering
    \includegraphics[width=0.47\textwidth]{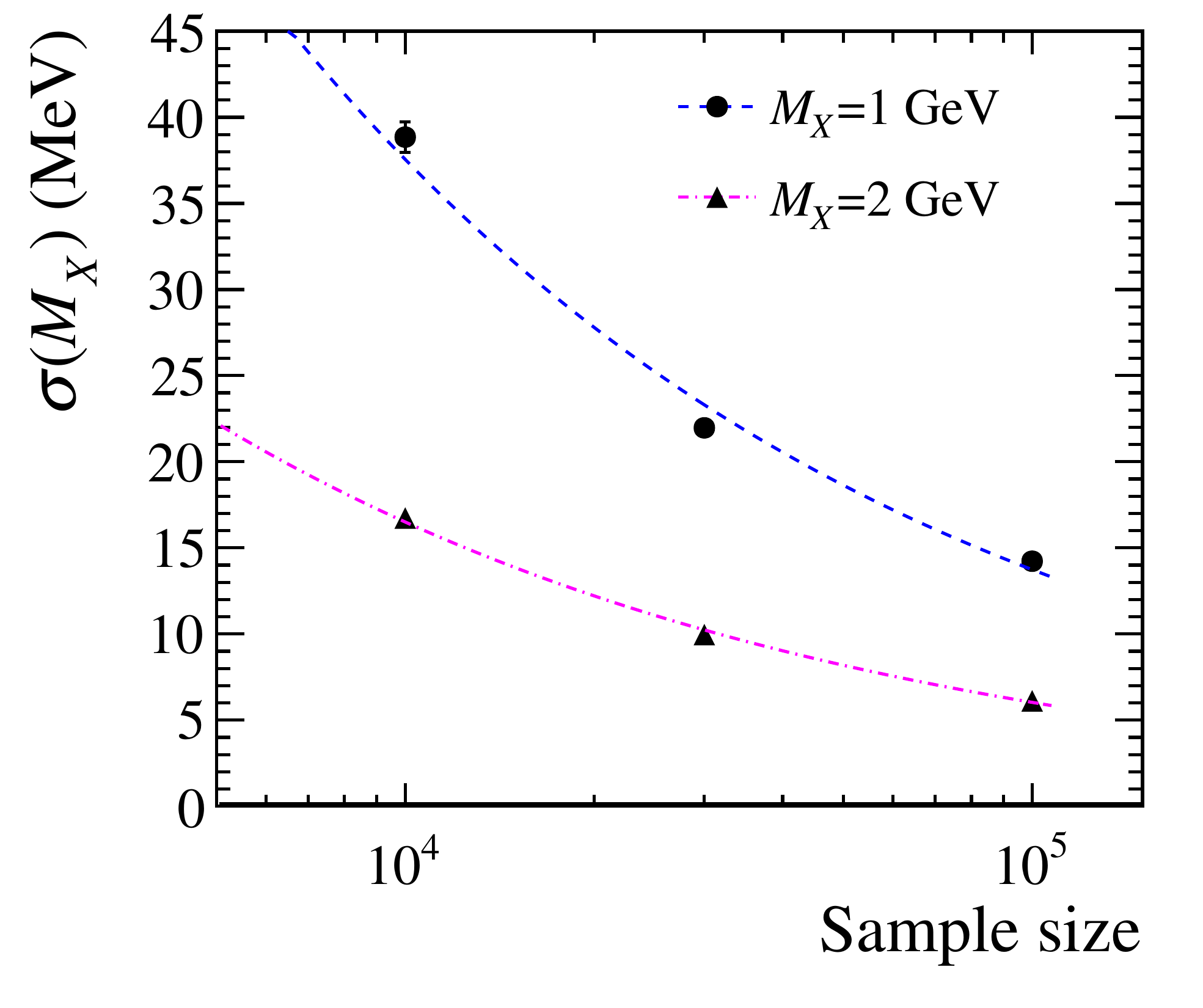}
    \put(-165, 144){(a)}
    \includegraphics[width=0.47\textwidth]{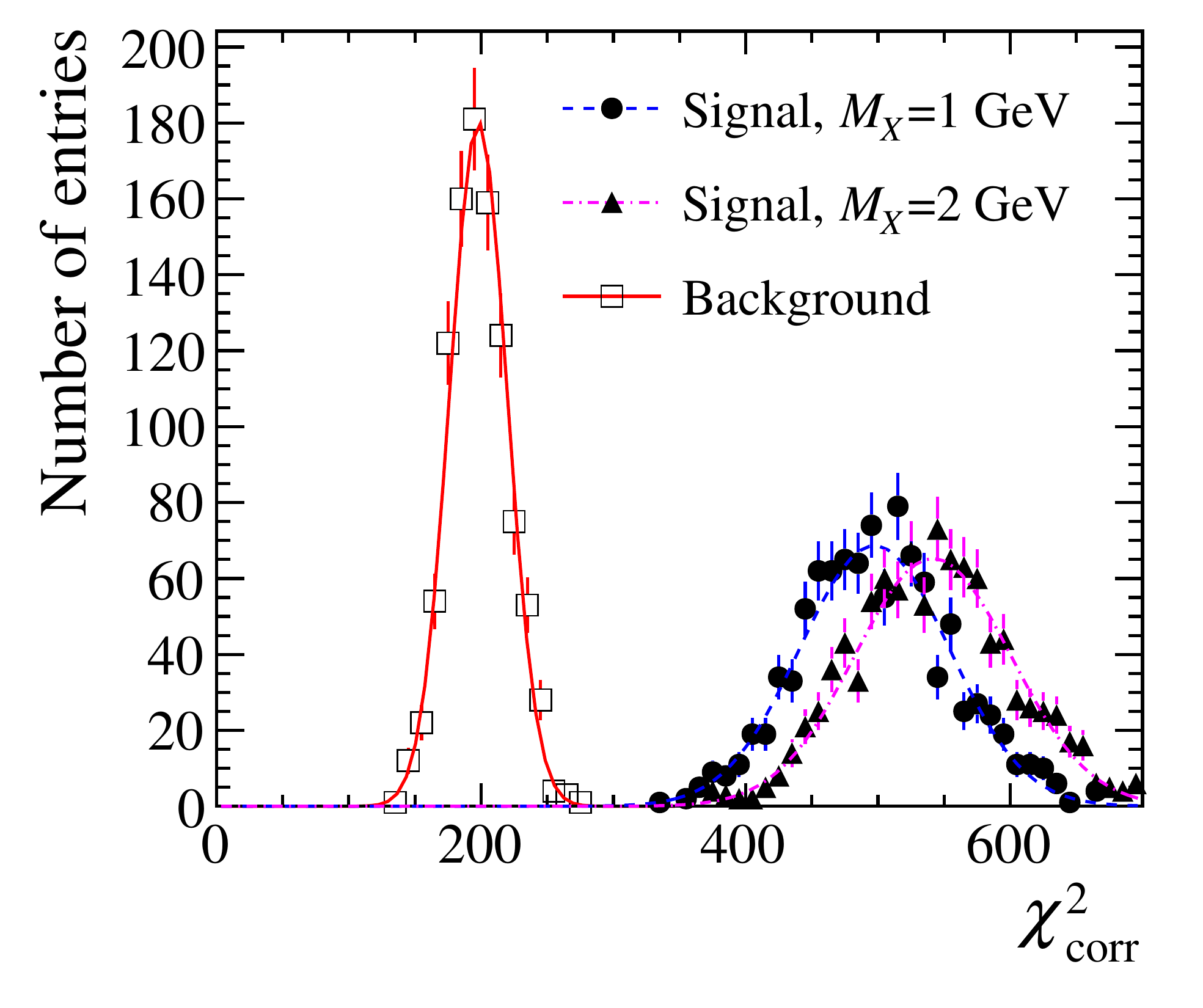}
    \put(-165, 144){(b)}
    \caption{(a) Mass resolutions for $M_X=1$\gev and $M_X=2$\gev 
            cases as a function of sample size. Smooth lines are the result of the fits with the function proportional to inverse square root of the sample size. 
    (b) Distributions of the test statistic $\chi^2_{\rm corr}$
    for background samples and signal with $M_X=1\gev$ and $M_X=2\gev$. Smooth lines are results of the Gaussian fits. }
    \label{fig:mass_res_chi2}
\end{figure}

The evaluation of mass resolution is performed with a simple approach, with the value $m_X$ at which the function $S_{\Bs}(m_X)$ crosses zero taken as the estimate of the measured mass $M^{\rm (meas)}_{X}$. Figures~\ref{fig:spec_dist}(b,d) show that this estimate gives a value that is somewhat biased towards higher masses (the true mass is indicated with a dashed line). The distributions of the measured masses $M^{\rm (meas)}_X$ for $M_X=1\gev$ and $M_X=2\gev$ and $10^4$ events of each $\Bs$ flavour are shown in Fig.~\ref{fig:mass_res}. The mass resolutions are $39.0\pm 1.1$\mev and $16.7\pm 0.4$\mev, and the means of the mass distributions are $1.0659\pm 0.0012$\gev and $2.0195\pm 0.0005$\gev, respectively. The mass resolution demonstrates a $1/\sqrt{N}$ scaling as a function of the sample size $N$ (see Fig.~\ref{fig:mass_res_chi2}(a)), while the mass bias is found to be independent of $N$. 

The bias in the measurement of mass using zero-crossing $S_{\Bs}(m_X)$ point originates from the not precisely sinusoidal behaviour of the flavour asymmetry due to reduction of coherence at higher decay times as a result of finite vertex resolution. An additional mass bias might be introduced in a real experiment by, \eg, non-uniform decay time acceptance. Such effects can be corrected using simulated data, either by introducing event-by-event weights to recover the sinusoidal behaviour of the flavour asymmetry, or by parametrising the bias and applying the correction to $M_X^{(\rm meas)}$. 

To illustrate the evaluation of the significance of the observed oscillation signal, a $\chi^2$-like test statistic accounting for correlation in the measured $C_{\Bs}$ and $S_{\Bs}$ values is constructed. Assuming that the contribution of peaking backgrounds is small, incoherent background with the same statistical properties (the uncertainties and the correlations of $C_{\Bs}$, $S_{\Bs}$ coefficients) as for the signal is taken as the null hypothesis. In that case, the test statistic that characterises the global significance of any $\Bs\to AX$ signal, with respect to the null hypothesis of no signal, is defined as 
\begin{equation}
    \chi^2_{\rm corr} = \sum\limits_{ij}\left( C_{\Bs}(m_i)\;\Sigma^{-1}_{C,ij}\; C_{\Bs}(m_j) + S_{\Bs}(m_i)\;\Sigma^{-1}_{S,ij}\; S_{\Bs}(m_j) \right), 
\end{equation}
where the terms $C_{\Bs}$ and $S_{\Bs}$ are measured at the values of mass $m_i$ (in our case, in 100 points uniformly distributed in the range from 0 to 3\gev), and $\Sigma^{-1}_{C,S}$ are the inverse covariance matrices for the $C_{\Bs}$ and $S_{\Bs}$ values at these masses:  
\begin{equation}
  \begin{split}
    \Sigma_{C,ij} & = \sqrt{\sigma_{C}(m_i), \sigma_{C}(m_j)}\; \rho_{C}(m_i, m_j), \\
    \Sigma_{S,ij} & = \sqrt{\sigma_{S}(m_i), \sigma_{S}(m_j)}\; \rho_{S}(m_i, m_j). 
  \end{split}
\end{equation}
The distributions of the test statistic $\chi^2_{\rm corr}$ for the signals and for the background are shown in Fig.~\ref{fig:mass_res_chi2}(b). The distributions for signal events and the background-only hypothesis are clearly separated, their difference provides a measure of the statistical significance of the effect. In this case, the significance of any deviation from the null hypothesis is around 10 standard deviations on average. In the real experiment, a similar test statistic can be used to assess the goodness of fit for a particular signal model, in which case one would need to calculate instead
\begin{equation}
    \chi^2_{\rm corr} = \sum\limits_{ij}\left( \Delta C_{\Bs}(m_i)\;\Sigma^{-1}_{C,ij}\; \Delta C_{\Bs}(m_j) + \Delta S_{\Bs}(m_i)\;\Sigma^{-1}_{S,ij}\; \Delta S_{\Bs}(m_j) \right), 
\end{equation}
where $\Delta C_{\Bs}$, $\Delta S_{\Bs}$ are the differences between the observed $C_{\Bs}$ and $S_{\Bs}$ functions and those expected from the model. 

\section{Conclusion}

A novel technique to study \Bs meson decays with invisible particles at a hadron collider has been presented. 
It uses the characteristic pattern of fast \Bs flavour oscillations to obtain information about the \Bs meson momentum and suppress any non-\Bs backgrounds. The technique has been studied 
with pseudoexperiments for a particular case of a decay $\Bs\to AX$, where the system $A$ is reconstructed 
and constrains the \Bs vertex, while $X$ is invisible. In this case, the proposed procedure provides sensitivity 
to the spectrum of $X$ masses which would be impossible with other approaches. 

The proposed technique can find its applications in the studies of both SM decays, 
\eg, involving neutrons, and for searches beyond the SM, such as decays with heavy 
neutrinos, axion-like particles and other dark matter candidates (see, \eg, Refs.~\cite{Elor:2018twp, Nelson:2019fln, Alonso-Alvarez:2019fym}). The benchmark measurements that should be feasible already with the currently available LHCb data set could employ 
Cabibbo-favoured open-charm decays such as $\Bs\to \Dsm \proton\overline{n}$. 
A similar approach is probably feasible for the reconstruction of decays with neutrinos (\eg, for the semileptonic
decays involving muons or $\tau$-leptons) where \Bs oscillations can provide additional kinematic constraints 
compared to conventional techniques used at LHCb. 

\section*{Acknowledgements}

The authors would like to thank Matthew Charles, Tim Gershon, Alexey Petrov, the CPPM LHCb group (Olivier Leroy, Julien Cogan) and members of the LHCb physics working group ``semileptonic decays'' (in particular, Greg Ciezarek, Lucia Grillo and Michel De Cian) for stimulating discussions and useful feedback on the paper. This work was supported by the A*MIDEX international project TAUFU, Aix-Marseille University.

\bibliographystyle{LHCb}
\bibliography{main}

\end{document}